\documentclass[aps,prc,letterpaper,preprintnumbers,superscriptaddress,showpacs,amsmath,letterpaper,floatfix,groupedaddress,nofootinbib,twocolumn]{revtex4}
\usepackage{hyperref}
\usepackage{multirow}
\usepackage{color}
\usepackage{tabularx}
\usepackage{graphicx}
\usepackage{amsthm}
\usepackage{amsmath}
\usepackage{amssymb}
\usepackage{bm}
\usepackage{textcomp}
\usepackage{url}
\usepackage{tabularx}
\usepackage{verbatim}

\graphicspath{ 
  {images/} 
}

\newcommand{\vz}{\mbox{$V_Z$}}
\newcommand{\pt}{\mbox{$p_T$}}
\newcommand{\bpt}{$\boldsymbol{\pt}$ }

\newcommand{\jpsi}{\mbox{$J/\psi$}}
\newcommand{\bjpsi}{$\boldsymbol{\jpsi}$ }

\newcommand{\auau}{\mbox{Au+Au}}

\newcommand{\bauau}{\mbox{\bf{Au+Au}}}
\newcommand{\cucu}{\mbox{Cu+Cu}}
\newcommand{\bcucu}{\mbox{\bf{Cu+Cu}}}

\newcommand{\pp}{\mbox{{\it p+p}}}

\newcommand{\snn}{\mbox{$\sqrt{s_{_{NN}}}$}}
\newcommand{\bsnn}{\mbox{$\boldsymbol{\snn}$} }

\newcommand{\fm}{\mbox{$\mathrm{fm}$}}
\newcommand{\cm}{\mbox{$\mathrm{cm}$}}

\newcommand{\gev}{\mbox{$\mathrm{GeV}$}}
\newcommand{\bgev}{\mbox{$\mathbf{GeV}$}}
\newcommand{\gevc}{\mbox{$\mathrm{GeV/}c$}}
\newcommand{\gevcc}{\mbox{$\mathrm{GeV/}c^2$}}
\newcommand{\mev}{\mbox{$\mathrm{MeV}$}}

\newcommand{\nsig}{n\sigma_{e}}

\newcommand{\nsigPi}{n\sigma_{\pi}}
\newcommand{\nsigP}{n\sigma_{p}}
\newcommand{\nsigK}{n\sigma_{K}}
\newcommand{\invbeta}{1/\beta}

\newcommand{\eop}{E/p}

\newcommand{\invbetacut}{|\invbeta-1|<0.03 }

\newcommand{\raa}{\mbox{$R_{\textit{AA}}$}}

\newcommand{\dedx}{\mbox{$\mathrm{d}{\it E}/\mathrm{d}{\it x}$}}

\newcommand{\ncoll}{\mbox{$N_{\textrm{coll}}$}}
\newcommand{\npart}{\mbox{$N_{\textrm{part}}$}}
\newcommand{\mncoll}{\mbox{$\langle N_{\textrm{coll}} \rangle$}}
\newcommand{\mnpart}{\mbox{$\langle N_{\textrm{part}} \rangle$}}
\newcommand{\impact}{\mbox{$b$}}
\newcommand{\mimpact}{\mbox{$\langle b \rangle$}}

\newcommand{\pA}{\it p+A}

\newcommand{\pps}{pp}
\newcommand{\AAa}{\mbox{\it A+A}}

\newcommand{\sigppjpsi}{\sigma_{\pps}}
\newcommand{\sigppinel}{\sigma_{\textrm{inel}}}

\begin{document}
\title{ $\bjpsi$ production at low $\bpt$ in $\bauau$ and $\bcucu$ collisions at\\ $\bsnn$ = 200 $\bgev$ at STAR }

\affiliation{AGH University of Science and Technology, Cracow, Poland}
\affiliation{Argonne National Laboratory, Argonne, Illinois 60439, USA}
\affiliation{University of Birmingham, Birmingham, United Kingdom}
\affiliation{Brookhaven National Laboratory, Upton, New York 11973, USA}
\affiliation{University of California, Berkeley, California 94720, USA}
\affiliation{University of California, Davis, California 95616, USA}
\affiliation{University of California, Los Angeles, California 90095, USA}
\affiliation{Universidade Estadual de Campinas, Sao Paulo, Brazil}
\affiliation{Central China Normal University (HZNU), Wuhan 430079, China}
\affiliation{University of Illinois at Chicago, Chicago, Illinois 60607, USA}
\affiliation{Cracow University of Technology, Cracow, Poland}
\affiliation{Creighton University, Omaha, Nebraska 68178, USA}
\affiliation{Czech Technical University in Prague, FNSPE, Prague, 115 19, Czech Republic}
\affiliation{Nuclear Physics Institute AS CR, 250 68 \v{R}e\v{z}/Prague, Czech Republic}
\affiliation{Frankfurt Institute for Advanced Studies FIAS, Germany}
\affiliation{Institute of Physics, Bhubaneswar 751005, India}
\affiliation{Indian Institute of Technology, Mumbai, India}
\affiliation{Indiana University, Bloomington, Indiana 47408, USA}
\affiliation{Alikhanov Institute for Theoretical and Experimental Physics, Moscow, Russia}
\affiliation{University of Jammu, Jammu 180001, India}
\affiliation{Joint Institute for Nuclear Research, Dubna, 141 980, Russia}
\affiliation{Kent State University, Kent, Ohio 44242, USA}
\affiliation{University of Kentucky, Lexington, Kentucky, 40506-0055, USA}
\affiliation{Korea Institute of Science and Technology Information, Daejeon, Korea}
\affiliation{Institute of Modern Physics, Lanzhou, China}
\affiliation{Lawrence Berkeley National Laboratory, Berkeley, California 94720, USA}
\affiliation{Massachusetts Institute of Technology, Cambridge, Massachusetts 02139-4307, USA}
\affiliation{Max-Planck-Institut f\"ur Physik, Munich, Germany}
\affiliation{Michigan State University, East Lansing, Michigan 48824, USA}
\affiliation{Moscow Engineering Physics Institute, Moscow Russia}
\affiliation{National Institute of Science Education and Research, Bhubaneswar 751005, India}
\affiliation{Ohio State University, Columbus, Ohio 43210, USA}
\affiliation{Old Dominion University, Norfolk, Virginia 23529, USA}
\affiliation{Institute of Nuclear Physics PAN, Cracow, Poland}
\affiliation{Panjab University, Chandigarh 160014, India}
\affiliation{Pennsylvania State University, University Park, Pennsylvania 16802, USA}
\affiliation{Institute of High Energy Physics, Protvino, Russia}
\affiliation{Purdue University, West Lafayette, Indiana 47907, USA}
\affiliation{Pusan National University, Pusan, Republic of Korea}
\affiliation{University of Rajasthan, Jaipur 302004, India}
\affiliation{Rice University, Houston, Texas 77251, USA}
\affiliation{University of Science and Technology of China, Hefei 230026, China}
\affiliation{Shandong University, Jinan, Shandong 250100, China}
\affiliation{Shanghai Institute of Applied Physics, Shanghai 201800, China}
\affiliation{SUBATECH, Nantes, France}
\affiliation{Temple University, Philadelphia, Pennsylvania 19122, USA}
\affiliation{Texas A\&M University, College Station, Texas 77843, USA}
\affiliation{University of Texas, Austin, Texas 78712, USA}
\affiliation{University of Houston, Houston, Texas 77204, USA}
\affiliation{Tsinghua University, Beijing 100084, China}
\affiliation{United States Naval Academy, Annapolis, Maryland, 21402, USA}
\affiliation{Valparaiso University, Valparaiso, Indiana 46383, USA}
\affiliation{Variable Energy Cyclotron Centre, Kolkata 700064, India}
\affiliation{Warsaw University of Technology, Warsaw, Poland}
\affiliation{University of Washington, Seattle, Washington 98195, USA}
\affiliation{Wayne State University, Detroit, Michigan 48201, USA}
\affiliation{Yale University, New Haven, Connecticut 06520, USA}
\affiliation{University of Zagreb, Zagreb, HR-10002, Croatia}

\author{L.~Adamczyk}\affiliation{AGH University of Science and Technology, Cracow, Poland}
\author{J.~K.~Adkins}\affiliation{University of Kentucky, Lexington, Kentucky, 40506-0055, USA}
\author{G.~Agakishiev}\affiliation{Joint Institute for Nuclear Research, Dubna, 141 980, Russia}
\author{M.~M.~Aggarwal}\affiliation{Panjab University, Chandigarh 160014, India}
\author{Z.~Ahammed}\affiliation{Variable Energy Cyclotron Centre, Kolkata 700064, India}
\author{I.~Alekseev}\affiliation{Alikhanov Institute for Theoretical and Experimental Physics, Moscow, Russia}
\author{J.~Alford}\affiliation{Kent State University, Kent, Ohio 44242, USA}
\author{C.~D.~Anson}\affiliation{Ohio State University, Columbus, Ohio 43210, USA}
\author{A.~Aparin}\affiliation{Joint Institute for Nuclear Research, Dubna, 141 980, Russia}
\author{D.~Arkhipkin}\affiliation{Brookhaven National Laboratory, Upton, New York 11973, USA}
% ...

\collaboration{STAR Collaboration}\noaffiliation

\preprint{Submitted to Phys. Rev. C}
\date{\today}
\pagenumbering{arabic}
\pagestyle{myheadings} 
\thanks{}
\markboth{{\small STAR Collaboration}}{{\small $\jpsi$ production in $\auau$ $\snn = 200 \ \gev$}} 



\begin{abstract}

The $\jpsi$ $\pt$ spectrum and nuclear modification factor ($\raa$) are reported for $\pt < 5 \ \gevc$ and $|y|<1$ from 0\% to 60\% central Au+Au and Cu+Cu collisions at $\snn = 200 \ \gev$ at STAR. A significant suppression of $\pt$-integrated $\jpsi$  production is observed in central Au+Au events. The Cu+Cu data are consistent with no suppression, although the precision is limited by the available statistics. $\raa$ in Au+Au collisions exhibits a strong suppression at low transverse momentum and gradually increases with $\pt$. The data are compared to high-$\pt$ STAR results and previously published BNL Relativistic Heavy Ion Collider results. Comparing with model calculations, it is found that the invariant yields at low $\pt$ are significantly above hydrodynamic flow predictions but are consistent with models that include color screening and regeneration.

\end{abstract}

\maketitle


\section{Introduction\label{introduction}}

Quantum chromodynamics (QCD) predicts a phase transition from hadronic matter to a partonic phase of matter, known as quark-gluon plasma (QGP), 
at high energy density and temperature. Ultrarelativistic heavy ion collisions provide a unique tool to create and study this strongly interacting matter that 
was thought to have populated 
the universe microseconds after the big bang. The production of heavy quarkonia has been extensively used to probe the medium created in heavy ion collisions, as these objects are expected to be suppressed in a deconfined medium owing to the Debye color screening of the heavy quark potential~\cite{ref:matsui,ref:qgp,ref:qgp2,ref:qgp3}. 
Because of their large mass, heavy quarks are primarily created in 
the initial hard scattering of the collision and thus provide information about the early stages and the evolution of the system. 
The production of the $c\bar{c}$ bound state-meson $\jpsi(1S)$ has been studied extensively at CERN Super Proton Synchrotron (SPS)~\cite{ref:na38,ref:na50,ref:na60}, BNL Relativistic Heavy Ion Collider (RHIC)~\cite{ref:phenixauau,ref:starauau}, 
and CERN Large Hadron Collider (LHC)~\cite{ref:lhc,ref:JpsiCMS}, and a $\jpsi$ suppression has been observed in heavy ion collisions. 

There are various modifications other than color screening to the production of $\jpsi$ in heavy ion collisions, 
such as the recombination of charm quarks~\cite{ref:recombination, ref:recombination2} into bound-state charmonium, 
and co-mover absorption~\cite{ref:comover, ref:comover2}. 
The formation time of the $\jpsi$ compared to the time required to emerge from the hot collisions volume may also allow for the escape of high transverse momentum ($\pt$) charmonium from the suppression region~\cite{ref:leakage} (so-called ``leakage'' effect). However, recent measurements of $\jpsi$ production at high-$\pt$ at RHIC~\cite{ref:starauau} show significant suppression in central Au+Au collisions at $\snn=200 \ \gev$ for $\pt>5 \ \gevc$. Also measurements at the LHC~\cite{ref:JpsiCMS} show a large suppression at high $\pt$, which suggests that there is only moderate leakage effect at RHIC and LHC energies. There are additional complications related to the feed-down from $B$-meson decays and excited states such as $\psi'$ and $\chi_{c}$. In $\pp$ collisions excited charmonia states contribute up to $40\%$ of the produced $\jpsi$ yield~\cite{ref:phenix:excited,ref:feed-down} while $B \rightarrow \jpsi$ yield depends strongly on $\pt$: it is $\sim 2\%$ at $\pt=1 \ \gevc$ and increases to 20\% for $\pt > 7 \ \gevc$~\cite{ref:starauau}. These sources will be modified in a hot medium and further influence the production in heavy ion collisions. There are also modifications from cold nuclear matter (CNM) effects~\cite{ref:cnm}, 
such as parton scattering~\cite{ref:partonscattering}, modifications to parton distribution functions (PDFs) inside the nucleus (shadowing)~\cite{ref:shadowing}, and nuclear absorption~\cite{ref:absorption}. To disentangle all of these effects a quantitative understanding of $\jpsi$ production in $\pp$, $\pA$, and $\AAa$ is required. The suppression owing to CNM effects has been intensively studied experimentally at Fermilab~\cite{Alde:1990wa,Leitch:1992di,Leitch:1999ea}, SPS~\cite{ref:cnm:sps,Alessandro:2004ap,Arnaldi:2010ky} and RHIC \cite{Adare:2012qf,ref:phenix:cnm:excited:states,Adare:2010fn} and a few significant effects were established (for instance an energy dependence of nuclear absorption and a large suppression of $\psi'$ in central $d+{\rm Au}$ collisions at RHIC~\cite{ref:phenix:cnm:excited:states}). However, a comprehensive understanding of the CNM effects is still missing. 

An important step towards understanding of $\jpsi$ in-medium interactions is a measurement of $\jpsi$ elliptic flow, which is sensitive to the production mechanism~\cite{ref:twocomp}. $\jpsi$ elliptic flow is consistent with zero for $\pt > 2 \ \gevc$~\cite{ref:JpsiV2}, indicating that $\jpsi$  is not produced dominantly by coalescence from thermalized (anti-) charm quarks in this $\pt$ range. The collision centrality and transverse momentum dependence of production rates in heavy ion collisions can provide further insight into the medium effects on $\jpsi$.
Recombination is expected to primarily populate low $\pt$ in central collisions where the charm quark density is the highest, 
while leakage effect and gluon scattering may enhance high-$\pt$ production.  
The comparison of production rates in different collision systems, such as $\auau$ and $\cucu$, can provide information about the system-size dependence of the modifications, as 
$\jpsi$ created in $\auau$ collisions will experience higher temperatures and a longer average path length through the surrounding nuclear matter. 

In this paper, the results for $\jpsi$ production in $\auau$ and $\cucu$ collisions at $\snn = 200 \ \gev$ at the STAR detector are reported. 
The $\jpsi$ $\pt$ spectrum and suppression at mid-rapidity ($|y|<1$) for $\pt < 5 \ \gevc$ in $0\--60\%$ centrality collisions are presented, and 
the transverse momentum and centrality dependence of the results are discussed. 
These results provide a set of complete spectra from one experiment to cover a wide range of 
transverse momentum and serve as a consistency check between different experiments in the overlapping kinematics and centralities.
This paper describes the experimental setup and data used in this analysis, followed by the analysis methods 
and associated efficiencies. The results are then discussed and compared to previous data and theoretical calculations. 

\section{Experiment and Data\label{experiment}}

The STAR experiment is a large-acceptance multi-purpose detector which covers a full azimuth and pseudorapidity range of $|\eta| < 1.8$~\cite{ref:stardetector}. 
The $\auau$ data used in this analysis were obtained using a minimum-bias trigger, which was defined as a coincidence signal in the east and west vertex position detectors (VPDs)~\cite{ref:vpd_det} located 5.7~m from the interaction point, in the pseudo-rapidity range of $4.2 \leq \eta \leq 5.1$. The VPD detector was not available in 2005 when $\cucu$ data were collected, and zero degree calorimeters (ZDCs)~\cite{ref:zdc_det} ($|\eta| > 6.3)$ were used instead in the minimum bias trigger. An additional trigger was used in $\auau$ collisions to identify central events ($0\--5\%$ most central collisions) by requiring a high occupancy in the time of flight (TOF) detector~\cite{ref:tof_det}.
The collision vertex position was determined using a Minuit vertex finder (MinuitVF)~\cite{ref:vertex_ppv}, 
and the vertex position along the beam line ($\vz$) was required to be within $30 \ \cm$ of the geometric center of STAR. This range was selected to maximize the uniformity of the detector acceptance. 
In the off-line analysis, a correlation between the $\vz$ measured in the VPD detector ($\vz^{\rm VPD}$ ) and the reconstructed collision vertex of $|\vz - \vz^{\rm VPD}| < 3 \ \cm$ 
was required to remove out-of-time (pile-up) events in $\auau$ collisions. 
A total of 27~M $\cucu$ and 189~M $\auau$ minimum-bias events, recorded in 2005 and 2010, respectively, 
  in $0\--60\%$ centrality collisions and satisfying the requirements described above were used in this analysis. 
An additional 85~M events in the $0\--5\%$ most central $\auau$ collisions recorded by the central trigger were also analyzed.  


\section{Analysis\label{analysis}}

$\jpsi$ reconstruction was performed via the dielectron decay channel, 
$\jpsi \rightarrow e^{+} + e^{-}$ with a branching ratio, $B$, of 5.9\%~\cite{ref:pdg}. The primary detector used for tracking and particle identification in this analysis is 
the time projection chamber (TPC). The barrel electromagnetic calorimeter (BEMC)~\cite{ref:bemc_det} and the TOF detector~\cite{ref:tof_det}, were used in the $\auau$ data analysis to improve the electron identification. 

The TPC is a large acceptance gas-filled detector and performs the tracking, momentum measurement and particle identification via the ionization energy loss ($\dedx$) of charged particles. The TPC has a full azimuthal coverage 
and a pseudorapidity coverage of $|\eta|< 1.8$. 

\begin {table} [tbp]
\begin {center}
\caption{The collision centrality definitions, average number of participants and binary collisions, and average impact parameter and their systematic uncertainties from the Glauber model~\cite{ref:glauber2} in $\auau$ collisions.}
\label{tab:auau_centrality}
\begin {ruledtabular}
\begin {tabular} { c c c c }
\noalign{\smallskip}
Centrality (\%) & $\npart$  & $\ncoll$ & $\impact \ (\fm)$ \\
\noalign{\smallskip}\hline\noalign{\smallskip}
             $ 0\--5 $ & $ 350 \pm 3 $ & $ 1071 \pm 29 $ & $ 2.3 \pm 0.1 $ \\
\noalign{\smallskip}
             $ 5\--10 $ & $ 300 \pm 7 $ & $ 856 \pm 27 $ & $ 4.0 \pm 0.2 $ \\
               \noalign{\smallskip}
             $ 10\--20 $ & $ 236 \pm 9 $ & $ 609 \pm 31 $ & $ 5.7 \pm 0.2 $ \\
               \noalign{\smallskip}
             $ 20\--30 $ & $ 168 \pm 11 $ & $ 377 \pm 33 $ & $ 7.3 \pm 0.3 $ \\
               \noalign{\smallskip}
             $ 30\--40 $ & $ 116 \pm 11 $ & $ 224 \pm 30 $ & $ 8.7 \pm 0.3 $ \\
               \noalign{\smallskip}
             $ 40\--50 $ & $ 76 \pm 11 $ & $ 124 \pm 25 $ & $ 9.9 \pm 0.4 $ \\
               \noalign{\smallskip}
             $ 50\--60 $ & $ 48 \pm 9 $ & $ 64 \pm 18 $ & $ 10.9 \pm 0.4 $ \\

                \noalign{\smallskip}\hline\noalign{\smallskip}
             $ 0-20 $ &  $ 280 \pm 6 $ & $ 785 \pm 29 $ & $ 4.4 \pm 0.2 $ \\
               \noalign{\smallskip}
             $ 20-40 $ &  $ 142 \pm 11 $ & $ 300 \pm 31 $ & $ 8.0 \pm 0.3 $ \\
               \noalign{\smallskip}
             $ 40-60 $ & $ 62 \pm 10 $ & $ 95 \pm 21 $ & $ 10.4 \pm 0.4 $ \\
                \noalign{\smallskip}\hline\noalign{\smallskip}
             $ 0-60 $ & $ 161 \pm 9 $ & $ 393 \pm 27 $ & $ 7.6 \pm 0.3 $ \\
\end {tabular}
\end {ruledtabular}
\end {center}
\end {table}

\begin {table} [tbp]
\begin {center}
\caption{The collision centrality definitions, average number of participants and binary collisions and their systematic uncertainties from the Glauber model~\cite{ref:glauber2} in $\cucu$ collisions~\cite{ref:phiauaucucu}.}
\label{tab:cucu_centrality}
\begin {ruledtabular}
\begin {tabular} { c c c c }
\noalign{\smallskip}
Centrality (\%) & $\npart$  & $\ncoll$  \\
		\noalign{\smallskip}\hline\noalign{\smallskip}
        $ 0-20 $ &  $ 87 \pm 1 $ & $ 156 \pm 12 $\\
        \noalign{\smallskip}
        $ 20-40 $ &  $ 46 \pm 1 $ & $ 63 \pm 4 $\\
        \noalign{\smallskip}
        $ 40-60 $ & $ 22 \pm 1 $ & $ 23 \pm 1 $ \\
        \noalign{\smallskip}\hline\noalign{\smallskip}
        $ 0-60 $ & $ 51 \pm 1 $ & $ 80 \pm 6 $\\
\end {tabular}
\end {ruledtabular}
\end {center}
\end {table}

The charged particle multiplicity was obtained from the number of reconstructed tracks in the TPC within $|\eta|<0.5$. 
Collision centrality was then determined from the measured multiplicity using a Glauber model~\cite{ref:glauber2}. 
For each collision centrality, an average impact parameter, $\mimpact$, average number of participants, $\mnpart$, and average number of binary collisions, $\mncoll$, were related to an observed multiplicity range. 
The centrality definitions in $\auau$ collisions are summarized in Table~\ref{tab:auau_centrality}, and 
the details on the centrality definitions for $\cucu$ can be found in \cite{ref:phiauaucucu} and Table ~\ref{tab:cucu_centrality}. 

We applied basic cuts to ensure good track quality. For $\auau$ data, we selected tracks with $|\eta|<1$, $\pt>0.2\ \gevc$, at least 16 points in the TPC and 52\% of the maximum number of possible TPC points. The distance of closest approach, $DCA$, to the collision vertex was required to be less than 2~cm. For $\cucu$ data, we used the same $\eta$ range but we required at least 25 points in the TPC, 55\% of the maximum number of possible TPC points and $DCA<1$~cm. We required electron candidates to have $\pt>1.1 \ \gevc$; it improved signal significance and did not affect the yield much because hadron rejection cuts described below removed the majority of electrons with $\pt<1.1 \ \gev$

\begin{figure}[h]
\begin{center}
\includegraphics[width=.5\textwidth]{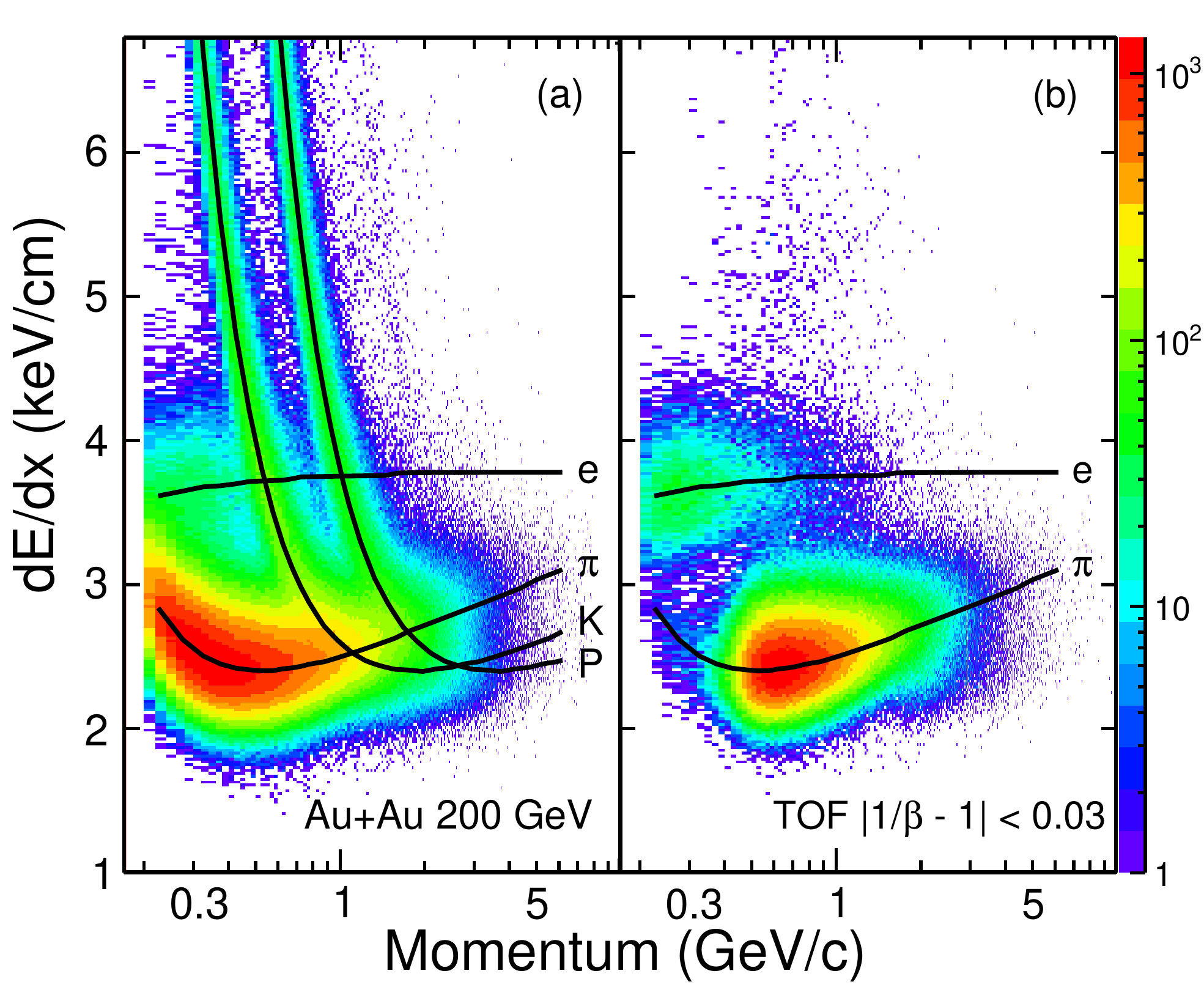}  
\caption{
  (Color online) The ionization energy loss $\dedx$ versus momentum in $\auau$ collisions for (a) all charged particles and (b) charged particles with $\invbetacut$. 
    The lines indicate the expected $\dedx$ for various particles obtained from the Bichsel functions~\cite{ref:bichsel}. 
}
\label{fig:dedx}
\end{center}
\end{figure}

The $\dedx$ distribution of charged particles in $\auau$ collisions is shown versus the momentum in Fig.~\ref{fig:dedx} (a). 
The expected $\dedx$ was obtained from Bichsel functions~\cite{ref:bichsel} and is shown for electrons, pions, kaons, and protons in Fig.~\ref{fig:dedx}. 
The measured $\dedx$ was normalized to the expected electron $\dedx$ to obtain $\nsig$, which is approximately Gaussian with $\mu=0$ and $\sigma=1$ for electrons:
\begin{equation}
\nsig = \ln\left(\frac{\dedx|_{\rm Measured}}{\dedx|_{\rm Bichsel}} \right)/\sigma_{\dedx} \ ,
\end{equation}
where $\dedx|_{\rm Measured}$ is the $\dedx$ measured by the TPC, $\dedx|_{\rm Bichsel}$ is the expected $\dedx$ for electrons obtained from the Bichsel functions, 
and $\sigma_{\dedx}$ is the $\dedx$ resolution. For the $\cucu$ analysis, electrons were required to satisfy $|\nsig|<2$.
The TOF detector was not available when the $\cucu$ data were taken. To remove contamination in the $\dedx$-crossover regions in the $\cucu$ analysis, hadrons were rejected using 
$|\nsigP| > 2.5$, $|\nsigK| > 2$, and $(\nsigPi> 2.5)$ or $(\nsigPi< -3)$, for protons, kaons, and pions, respectively. 
In $\auau$ collisions, electrons were identified with the TPC by requiring $-1 < \nsig < 2$, and hadrons were further rejected using the TOF and BEMC, as described below.

\begin{figure}[h]
\begin{center}
\includegraphics[width=.5\textwidth]{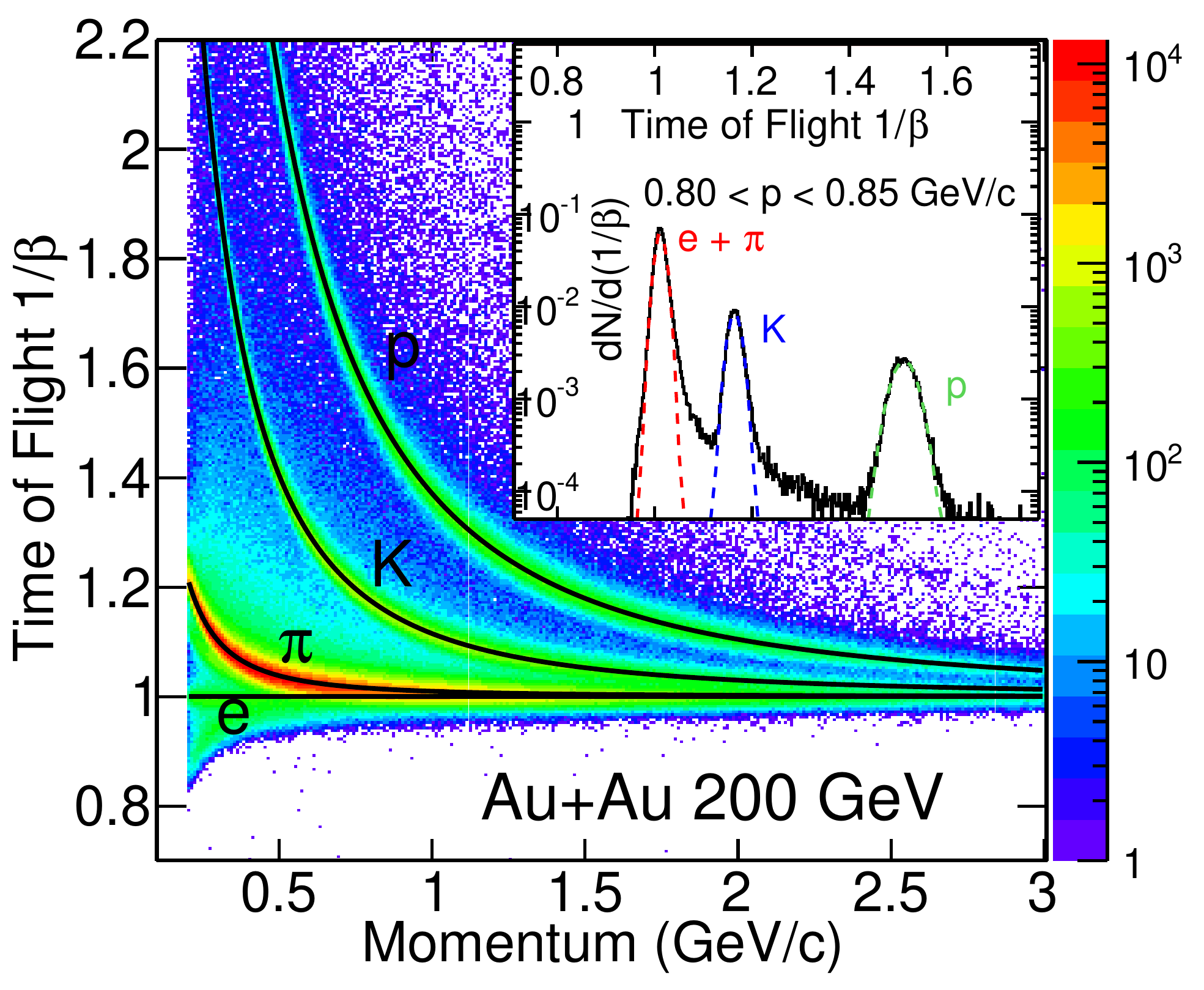} 
\caption{
  (Color online) The TOF $\invbeta$ versus momentum for charged particles in $\auau$ collisions. 
    The lines indicate the expected $\invbeta$ for various particles. Inset is the $\invbeta$ distribution for $0.8 < p < 0.85 \ \gevc$.
}
\label{fig:tof}
\end{center}
\end{figure}

The TOF was used in the $\auau$ data analysis to improve the electron-hadron discrimination, especially where the electron and hadron $\dedx$ values overlap. 
The TOF measures the velocity of charged particles. 
Electrons were identified by selecting fast particles, which was done by requiring 
$|1/\beta-1|<0.03$, where $\beta = v/c$~\cite{ref:tof_nim}. The distribution of $\invbeta$ versus momentum for all charged particles is shown in Fig.~\ref{fig:tof}, 
and the expected values for electrons, pions, kaons, and protons are also indicated. 
The inset in the diagram is the $\invbeta$ distribution for $0.8 < p < 0.85 \ \gevc$. The electrons and pions are clearly separated from the heavier hadrons such as kaons and protons. Owing to finite time resolution of the TOF, we observe a small number of tracks with $1/\beta < 1$. 
The TPC has a limited capacity for separation of electrons from kaons and protons below $\sim 1 \ \gevc$. The TOF extends the electron identification 
capabilities to low $\pt$ by separating electrons and heavier hadrons for $p < 1.5 \ \gevc$.  
The $\dedx$ distribution for charged particles in $\auau$ collisions is shown in Fig~\ref{fig:dedx} (a) before using the TOF and Fig~\ref{fig:dedx} (b) after requiring $\invbetacut$. After using the TOF, the heavier hadrons are removed and the electron band is separated from the remaining hadrons. 
Pions, which are too light to effectively separate from electrons using the TOF, as seen in Fig~\ref{fig:dedx} (b), are removed using $\dedx$ by requiring $-1 < \nsig < 2$. 
At high momentum, the TOF is no longer effective at separating electrons and hadrons. For $p > 1.5 \ \gevc$, the BEMC is used to improve the electron identification, as described below. 

\begin{figure}[h]
\begin{center}
\includegraphics[width=.5\textwidth]{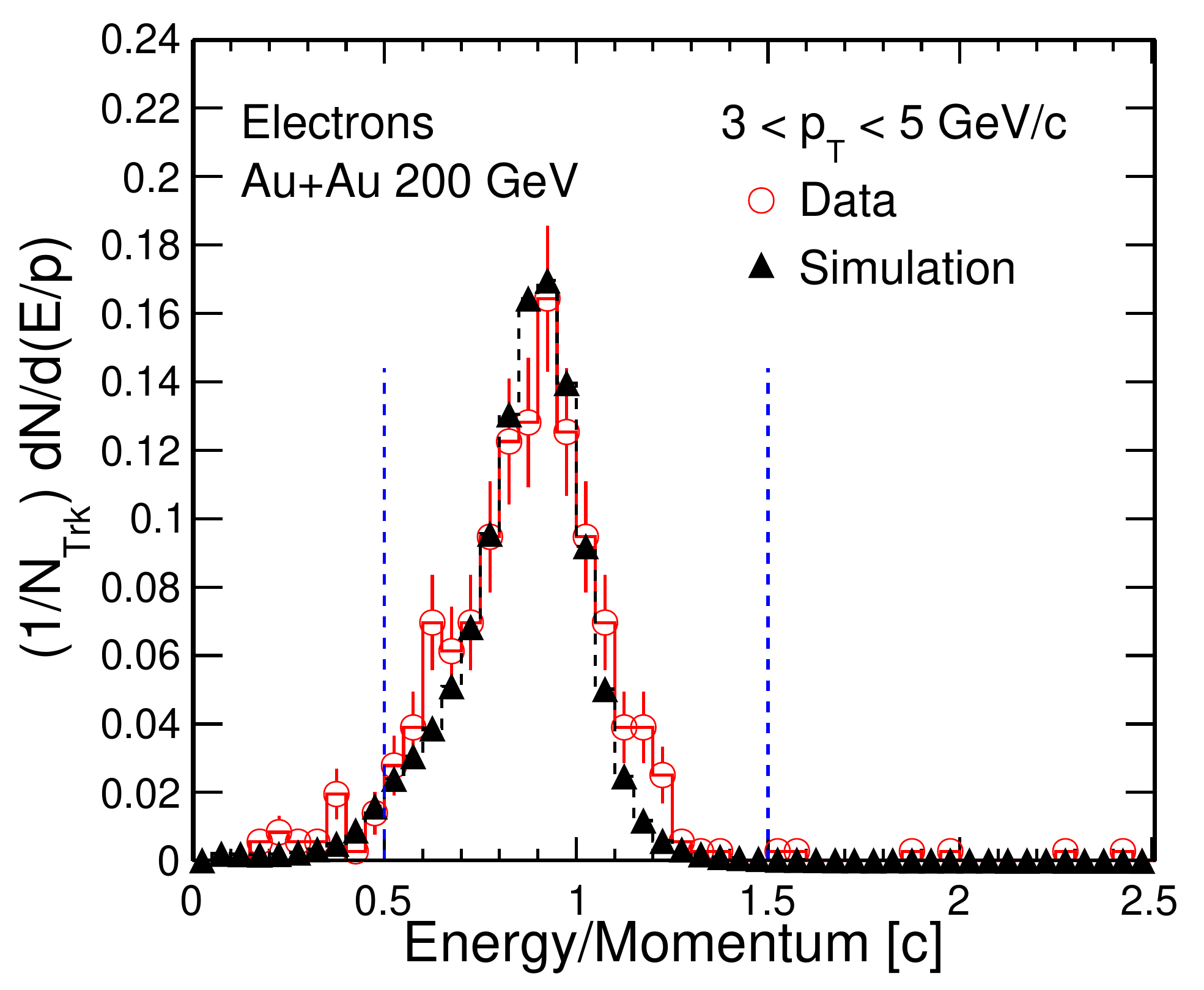} 
\caption{
  (Color online) The $\eop$ ratio for $3 < \pt < 5 \ \gevc$, where $E$ is the single-tower energy from the BEMC, and $p$ is the momentum from the TPC. 
    A high-purity ($>95\%$) electron sample from data (open circles) is compared to 
    a GEANT simulation of the $\eop$ for Monte Carlo electrons (solid triangles). 
    The dashed vertical lines indicate the accepted region ($0.5 < \eop < 1.5$).
}
\label{fig:eop}
\end{center}
\end{figure}

The BEMC is a lead-scintillator calorimeter segmented into 4800 towers with a tower size of $\Delta\eta\times\Delta\phi = 0.05\times 0.05$.  
This detector has a total radiation length of $\sim 20X_{0}$ and achieves an energy resolution of $\mathrm{d}E/E \sim 16\%/\sqrt{E}$~\cite{ref:bemc_det}. 
The BEMC contains a barrel shower maximum detector at a radiation length of $\sim5X_{0}$ which consists of two layers of gas wire pad chambers along 
the $\eta$ and $\phi$ 
planes. It was used to determine the position of energy deposits in the BEMC. 

For $p > 1.5 \ \gevc$, the BEMC was used to separate electrons from hadrons in $\auau$ collisions using the 
energy-to-momentum ratio, $\eop$, where $p$ is the momentum obtained from the TPC and $E$ the single-tower energy obtained from the BEMC. 
The energy-to-momentum ratio is shown in Fig.~\ref{fig:eop} for a high purity ($>95\%$) electron sample from data, which was obtained by using $\dedx$, TOF and selecting photonic electrons (from photon conversion in the detector material or from Dalitz decays of $\pi$ and $\eta$ mesons).

Comparison of the measured electron $\eop$ to that for Monte Carlo electrons from a full GEANT simulation~\cite{ref:p_pi_dAu} shows good agreement.
There is a non-Gaussian tail at low $\eop$ seen in real and simulated data owing to energy loss in neighboring towers when an electron strikes near the tower edge. 
The BEMC was used to discriminate electrons and hadrons by requiring $0.5 < \eop < 1.5$.  

\section{Signal and Corrections\label{signal}}

\begin{figure}[hbt!!!]
\begin{center}
\includegraphics[width=.5\textwidth]{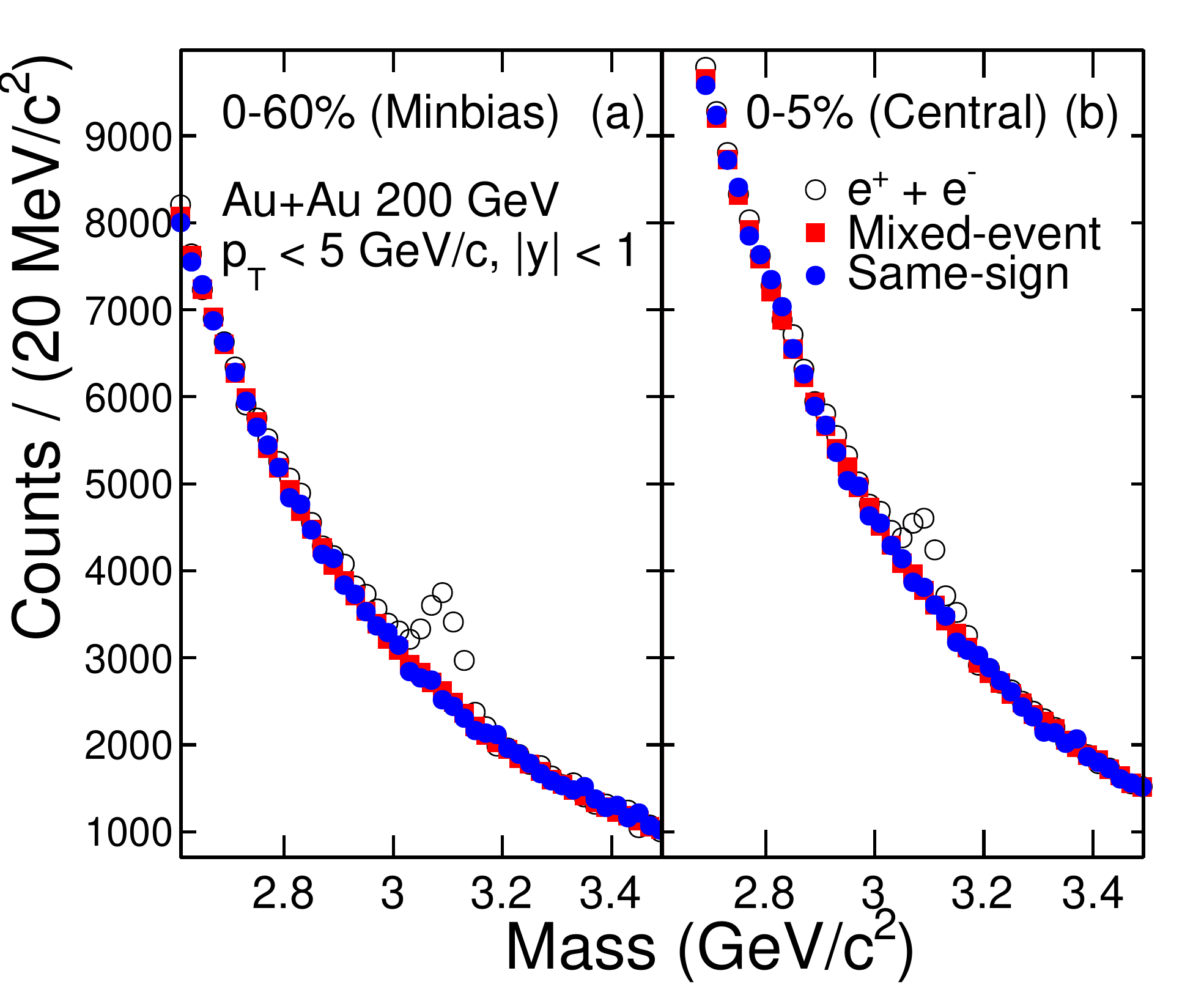}   
\caption{
  (Color online) The opposite-sign dielectron invariant mass distribution (open circles) from (a) minimum bias trigger data in $0\--60\%$ 
    and (b) central trigger data in $0\--5\%$ centrality $\auau$ collisions at $\snn = 200 \ \gev$. The mixed-event background (squares) was 
    normalized to the like-sign background (open circles) and subtracted from the opposite-sign distribution to obtain the $\jpsi$ signal.
}
\label{fig:mass}
\end{center}
\end{figure}

\begin{figure}[hbt!!!]
\begin{center}
\includegraphics[width=.5\textwidth]{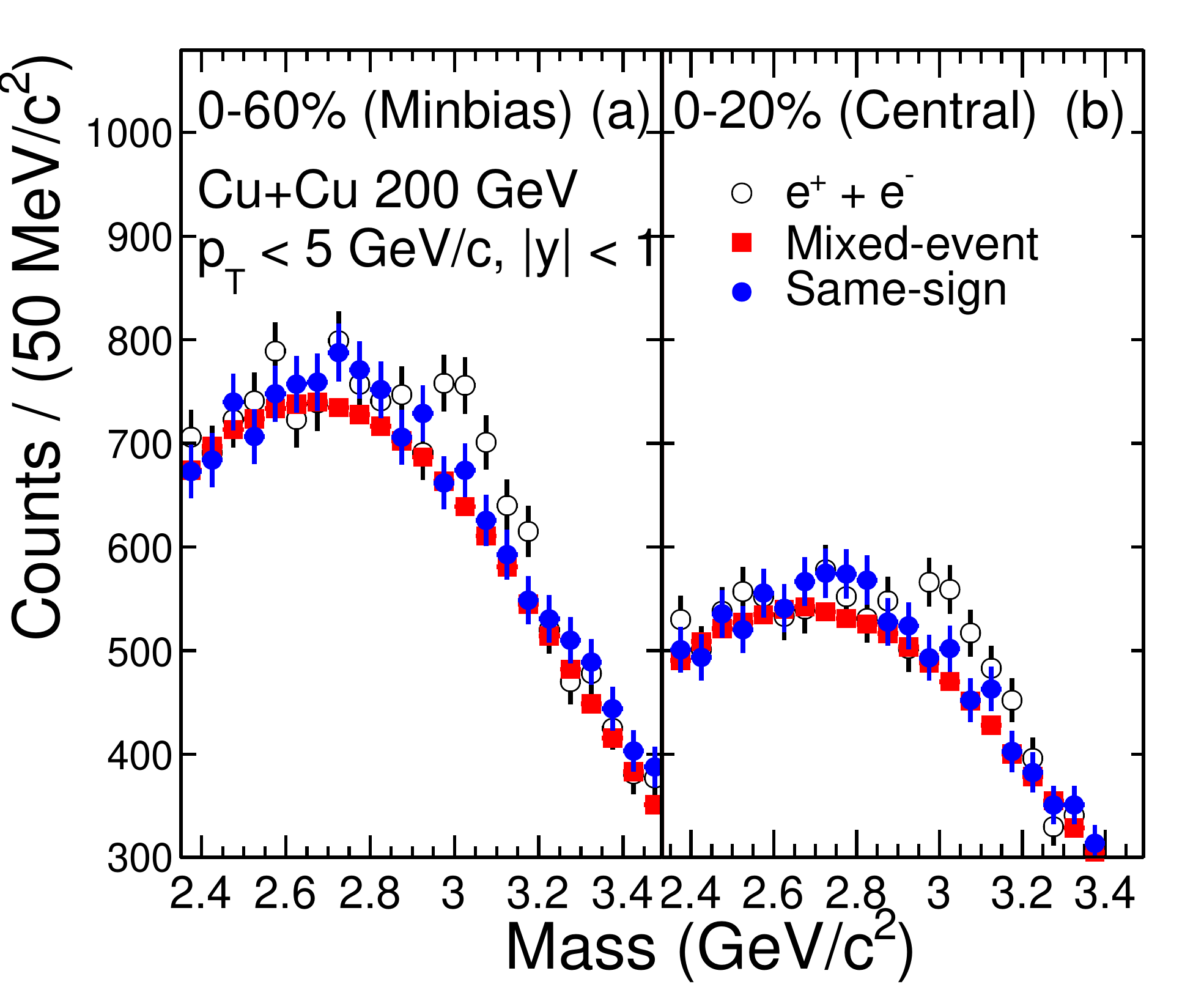}    
\caption{
  (Color online) The opposite-sign dielectron invariant mass distribution (open circles) from (a) minimum bias ($0\--60\%$ centrality) and (b) central ($0\--20\%$ centrality) $\cucu$ collisions at $\snn = 200 \ \gev$. The mixed-event background (squares) was normalized to the like-sign background (open circles) and subtracted from the opposite-sign distribution to obtain the $\jpsi$ signal.
}
\label{fig:mass:cucu}
\end{center}
\end{figure}

The opposite-sign dielectron invariant mass spectrum is shown for $\auau$ collisions with $\pt < 5 \ \gevc$ and $|y|<1$ in Fig.~\ref{fig:mass}, and was obtained from 
(a) minimum bias data in $0\--60\%$ centrality and (b) central-triggered data in $0\--5\%$ centrality. Figure \ref{fig:mass:cucu} shows similar distributions for $\cucu$ collisions for minimum-bias ($0\--60\%$) and central ($0\--20\%$) collisions. The combinatorial background was estimated using 
same-sign pairs from the same event, and opposite-sign pairs from mixed events. 
The opposite-sign mixed-event background was normalized to the same-sign same-event background in a mass range of $2.6 < m_{ee} < 3.6 \ \gevcc$ and subtracted from the dielectron invariant mass spectrum to obtain the $\jpsi$ signal. An effect of possible different acceptance for same-sign and opposite-sign pairs was studied for the same data set in Ref.~\cite{ref:star:diele:v2}. The ratio of same-sign and opposite-sign pairs is unity for $m_{ee} > 0.5 \ \gevcc$, thus the impact on the mixed-event background normalization is negligible.

A signal-to-background ratio of $1:20$ was achieved in $0\--60\%$ centrality $\cucu$ collisions. 
This has substantially improved with the removal of $\gamma$-converting material of the inner-detector subsystems and the addition of the TOF. A signal-to-background ratio of $1:5$ in $0\--60\%$ centrality $\auau$ collisions was achieved, increasing 
from $1:11$ in $0\--5\%$ to $1:1$ in $40\--60\%$ centrality collisions.  

\begin{figure}[hbt!!!]
\begin{center}
\includegraphics[width=.5\textwidth]{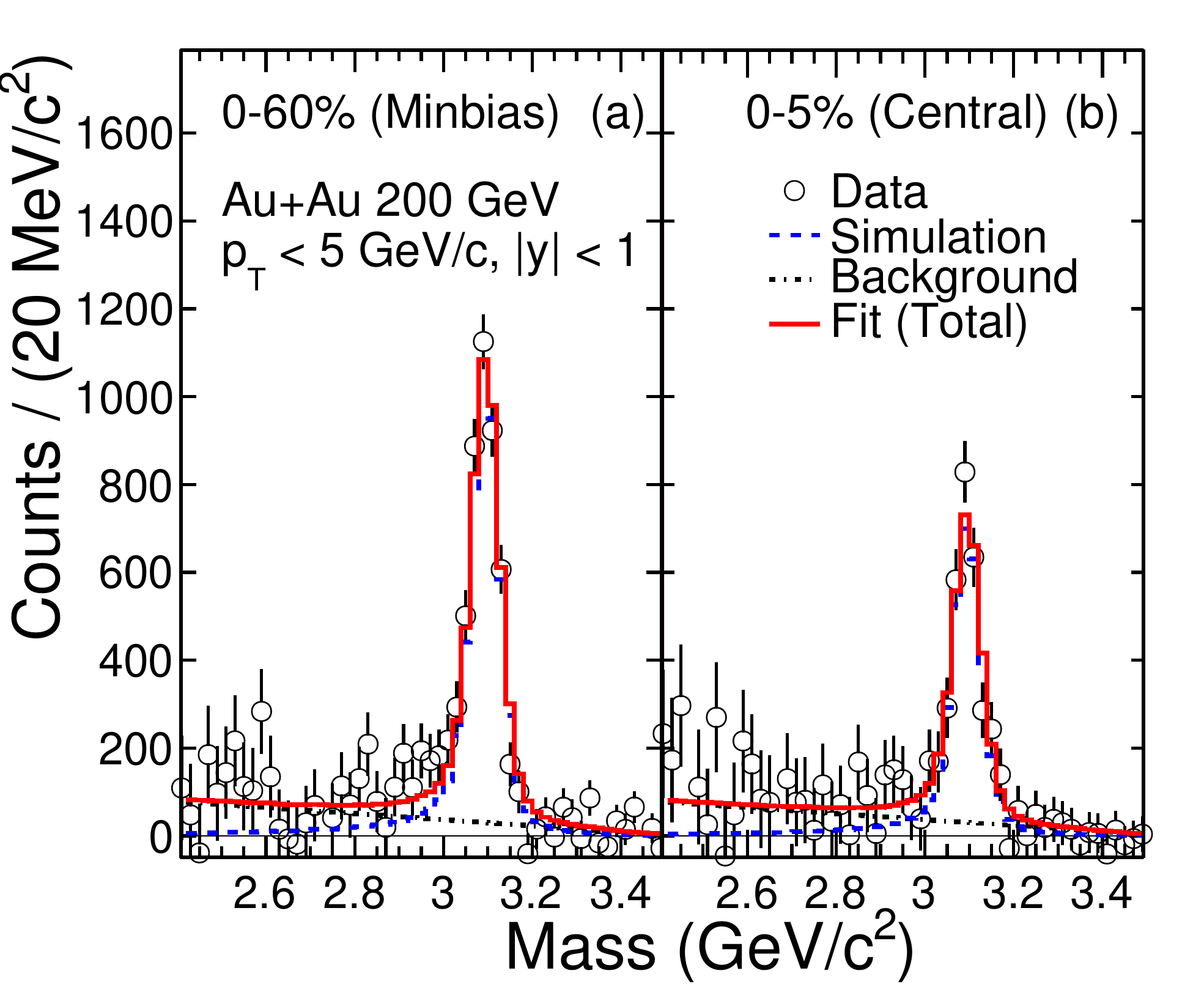}     
\caption{
  (Color online) The $\jpsi$ signal for $|y|<1$ after mixed-event background subtraction (open circles) 
    from (a) minimum bias trigger data in $0\--60\%$ centrality and (b) central trigger data in $0\--5\%$ centrality $\auau$ collisions at $\snn = 200 \ \gev$.  
    The $\jpsi$ signal shape obtained from a simulation (dashed line) is combined with a linear background (dot-dashed line) and is fitted to the data (solid line).
}
\label{fig:signal}
\end{center}
\end{figure}

\begin{figure}[hbt!!!]
\begin{center}
\includegraphics[width=.5\textwidth]{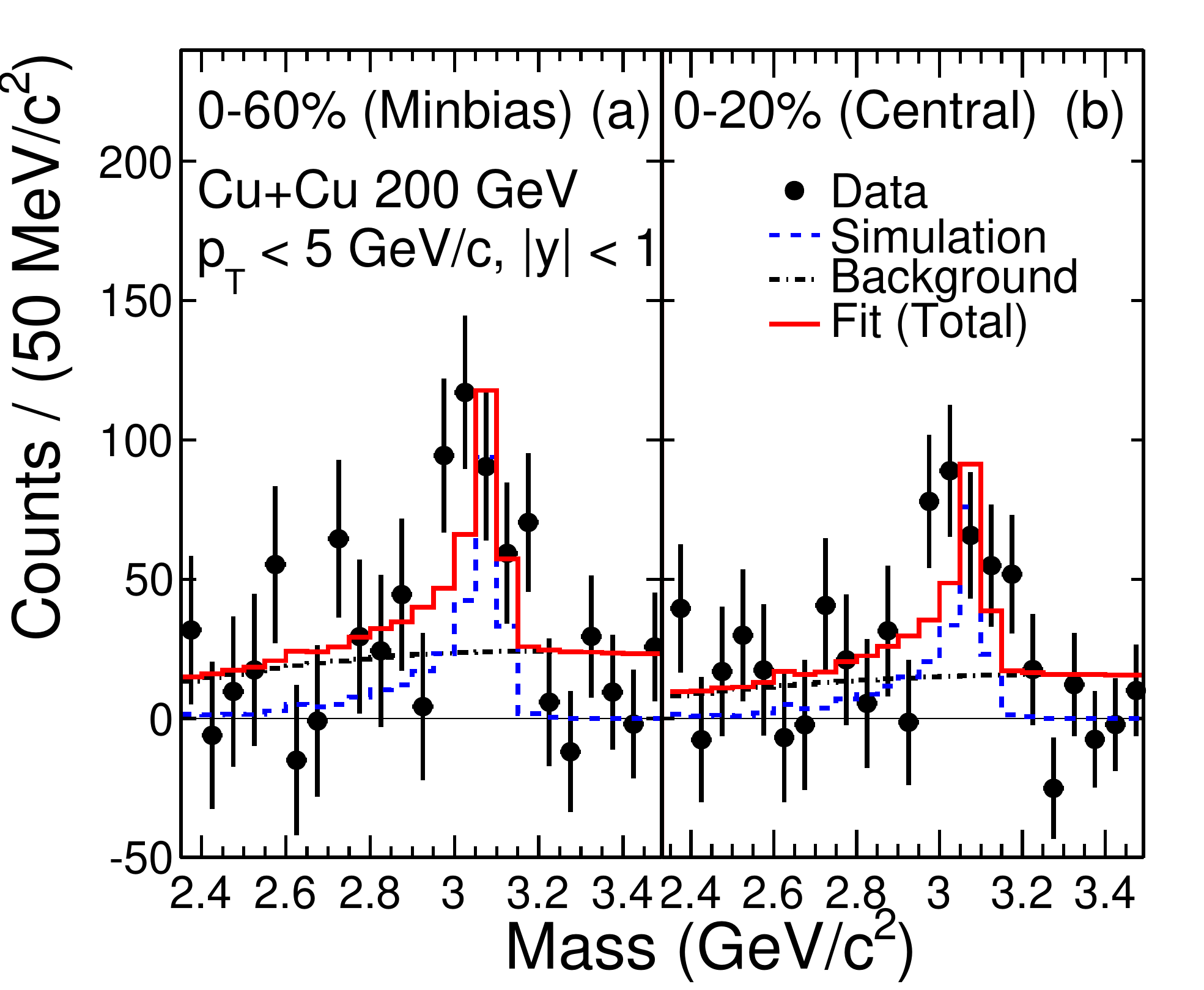}     
\caption{
  (Color online) The $\jpsi$ signal for $|y|<1$ after mixed-event background subtraction (open circles) 
     from minimum bias ($0\--60\%$ centrality) and (b) central ($0\--20\%$ centrality) $\cucu$ collisions at $\snn = 200 \ \gev$.  
    The $\jpsi$ signal shape obtained from a simulation (dashed line) is combined with a second-order polynomial background (dot-dashed line) and is fitted to the data (solid line).
}
\label{fig:signal:cucu}
\end{center}
\end{figure}

The dielectron invariant mass spectrum after background subtraction in $\auau$ and  $\cucu$  collisions is shown in Fig.~\ref{fig:signal} and Fig.~\ref{fig:signal:cucu}. The data are compared to the $\jpsi$ signal shape obtained from a simulation, combined with a straight line (in the case of $\auau$) or second-order polynomial (for $\cucu$) background. The $\jpsi$ signal shape was determined using a GEANT simulation of the detector response to Monte Carlo $\jpsi$ particles embedded into real data events, and is due to the resolution of the TPC and bremsstrahlung of the daughter electrons in the detector. 

The yield was calculated by counting the entries in a mass window of $2.7 < m_{ee} < 3.2 \ \gevcc$ as a function of collision centrality and transverse momentum. To account for residual background in $\auau$ collisions, a straight line was included in the  $\jpsi$ signal shape fit to the $\jpsi$ signal in the data. In $\cucu$ collisions, the simulated $\jpsi$ signal shape does not reproduce the data well and we used a Gaussian instead. Furthermore, residual background had a different shape owing to the $p_T$ cut for electrons (especially for low-$p_T$ $\jpsi$), thus a second-order polynomial was used to estimate this background. The residual background was then subtracted from the counts in the given mass range. 

The fraction of $\jpsi$ counts outside of the mass range of $2.7 < m_{ee} < 3.2 \ \gevcc$ was determined from the $\jpsi$ signal shape obtained from simulation, and was found to be $7-8\%$ for both $\auau$ and $\cucu$ collisions. This was used to correct the number of $\jpsi$ counts. 
A total of $370\pm90$ $\jpsi$ were reconstructed in $\cucu$ collisions with signal significance $S/\Delta S = 4$, 
   where $\Delta S$ is the uncertainty on the measured signal $S$. 
   In $\auau$, $5636 \pm 295$ $\jpsi$ were reconstructed in minimum bias $0\--60\%$ collisions with a significance of $19$, 
while $4050\pm 322$ $\jpsi$ were reconstructed in central-trigger $0\--5\%$ centrality collisions with a significance of $13$.  

\begin{figure}[h]
\begin{center}
\includegraphics[width=.5\textwidth]{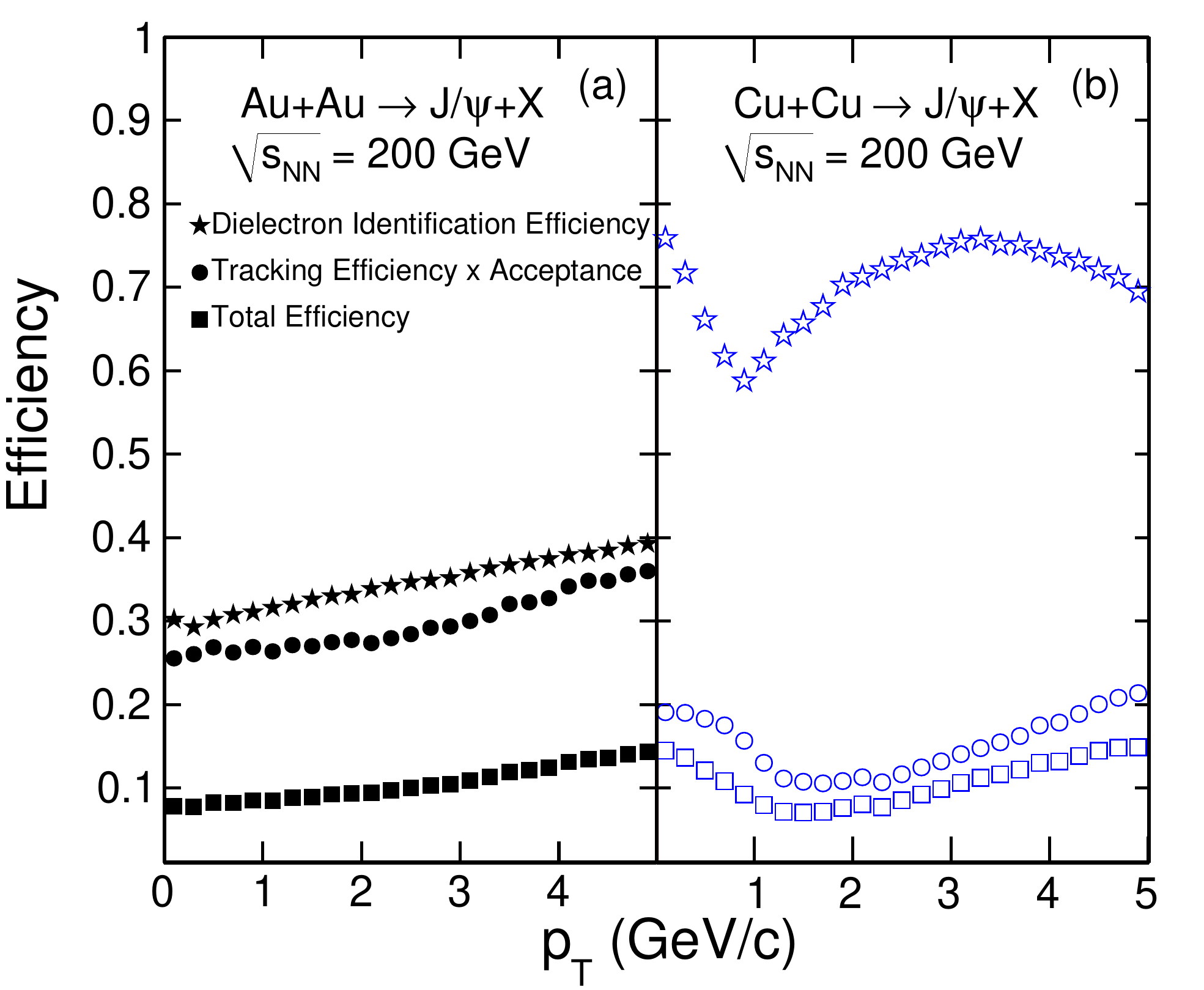}     
\caption{
  (Color online) The $\jpsi$ total efficiency (squares), tracking efficiency and acceptance (circles), and dielectron identification efficiency (stars) 
    for 0\--60\% centrality (a) $\auau$ (solid symbols) and (b) $\cucu$ (open symbols) collisions. 
      The shape of the dielectron identification efficiency in (b) is attributable to a proton rejection cut (see text for details).  
}
\label{fig:efficiency}
\end{center}
\end{figure}

The detector acceptance and tracking efficiency were determined using a Monte Carlo GEANT simulation, 
and are shown in Fig.~\ref{fig:efficiency} for $0\--60\%$ centrality collisions in (a)~$\auau$ and (b)~$\cucu$. 
The tracking efficiency is higher in $\auau$ compared to $\cucu$ data due to stringent $\pt$ and track quality cuts for electron candidates. The electron identification efficiency in $\cucu$ exhibits a strong $\pt$ dependence owing to the hadron rejection requirements placed on $\dedx$, which 
cause a drop in the identification efficiency for $\pt \sim 1 \ \gevc$. 
The identification efficiency is lower in $\auau$ due to the TOF matching efficiency of $\sim 65\%$ and BEMC matching efficiency of $\sim 85\%$. The TOF and BEMC matching efficiencies were calculated from the ratio of all electron candidates to those which were successfully matched to the TOF and BEMC, respectively.  
The total $\jpsi$ efficiency and acceptance correction is obtained by combining the tracking efficiency and acceptance with the 
total identification efficiency of both electron daughters, and is $\sim 8-15\%$ ($5-15\%$) in $\auau$ ($\cucu$) collisions. 

The systematic uncertainties in the $\auau$ analysis include uncertainties from the particle identification efficiency using the TPC ($6\%$), TOF ($3\%$), and BEMC ($15\%$),  
    the tracking efficiency and acceptance correction ($7\%$), and the yield extraction methods ($10\%$). 
    The uncertainty on the yield extraction was determined by varying the width of the $\jpsi$ signal shape from simulation, 
    by varying the mass range in which the fit was performed, and by comparing the yields obtained from fitting and from counting. 
    An additional $4\%$ uncertainty was included to account for the contribution from radiative decay $\jpsi \rightarrow e^+ + e^- + \gamma$~\cite{ref:intern:rad}, which are not included in the simulation. 
In the case of the $\cucu$ analysis, the main sources of systematic uncertainty were from the estimation of the combinatorial background ($\sim 13-26\%$), and the tracking efficiency and acceptance correction ($4\%$). The particle identification was performed using the TPC only, and resulted in an uncertainty of $\sim3\%$. The systematic uncertainties in the $\auau$ and $\cucu$ data for $0\--60\%$ centrality and integrated $\pt$ are summarized in Table~\ref{tab:systematics}. The centrality and transverse momentum 
 dependence of the total systematic uncertainties, quoted as one standard deviation, are reflected in the results shown in Section~\ref{results}.

\begin {table} [tbp]
\begin {center}
\caption{The systematic uncertainties for $0\--60\%$ centrality in $\auau$ and $\cucu$ collisions. } 
\label{tab:systematics}
\begin{tabularx}{0.48\textwidth}{X X X}
\noalign{\smallskip}
\hline\hline
\noalign{\smallskip}
\multirow{2}{*}{Source} & \multicolumn{2}{l} {Relative Uncertainty (\%)} \\
                        & $\cucu$ & $\auau$ \\
\hline\noalign{\smallskip}

eID (TPC) & $3$ & $6$ \\
eID (TOF) & - & $3$ \\
eID (BEMC) & - & $14$ \\
Efficiency & $4$ & $7$ \\
Yield & $13$ & $10$ \\
Total & $14$ & $21$ \\
\hline\noalign{\smallskip}
$\ncoll$ & $7$ & $7$ \\
$\sigppinel$ & {$8$} & {$8$}\\
$\sigppjpsi \ $  {\small(stat.)} & {$3$} & {$3$}\\
$\sigppjpsi \ $  {\small(syst.)} & {$7$} & {$7$}\\

\noalign{\smallskip}
\hline\hline

\end{tabularx}
\end {center}
\end {table}

\section{Results\label{results}}

\begin{figure*}[h]
\begin{center}
\includegraphics[width=.7\textwidth]{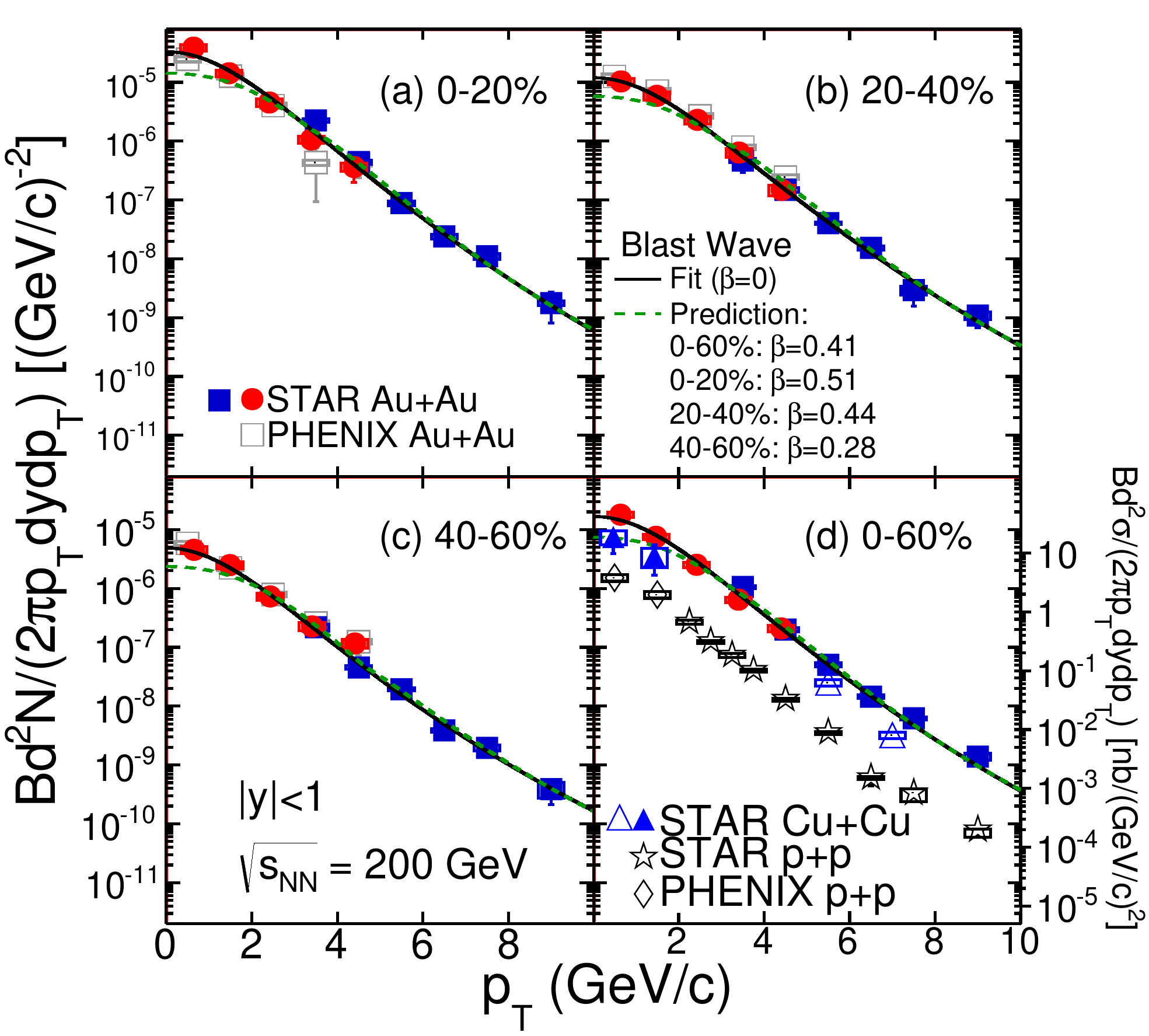}    
\caption{
  (Color online) The invariant yield versus transverse momentum for $|y|<1$ in (a) $0\--20\%$, (b) $20\--40\%$, (c) $40\--60\%$, and (d) $0\--60\%$ 
    centrality in $\auau$ collisions (solid circles).  
    The results are compared to high-$\pt$ ($3 < \pt < 10 \ \gevc$) results from STAR~\cite{ref:starauau} (solid squares) and PHENIX data~\cite{ref:phenixauau} (open squares). Also shown is the 
    yield in 0\--60\% centrality $\cucu$ collisions for low $\pt$ (solid triangles) and high $\pt$~\cite{ref:starpp} (open triangles). 
    The models are described in the text~\cite{ref:startbw1, ref:startbw2}.
    The $\jpsi$ cross section in $\pp$ collisions  is also shown in (d) at STAR (stars) and PHENIX~\cite{ref:JpsiPpData} (diamonds), and the scale is indicated on the right axis. 
}
\label{fig:ptspectrum}
\end{center}
\end{figure*}

The $\jpsi$ invariant yield is defined as
\begin{equation}
\frac{B}{2\pi\pt}\frac{\mathrm{d}^2N}{\mathrm{d}y\mathrm{d}\pt} = \frac{1}{2\pi\pt\Delta\pt\Delta y}\frac{N_{\small{J/\psi}}}{N_{\mathrm{Ev}}\varepsilon},
\end{equation}
where $N_{\small{J/\psi}}$ is the uncorrected number of reconstructed $\jpsi$, $B$ is branching ratio, $N_{\mathrm{Ev}}$ is the number of events, and $\varepsilon$ is the 
total efficiency and acceptance correction factor. The $\pt$-dependence of the $\jpsi$ invariant yield from this analysis is shown in Fig.~\ref{fig:ptspectrum} (d) for 
$\pt < 5 \ \gevc$ and $|y|< 1$ in $0\--60\%$ centrality $\auau$ and $\cucu$ collisions and compared to the
high-$\pt$ $\jpsi$ yield for $\cucu$ collisions~\cite{ref:starpp} (for $5 < \pt < 8 \ \gevc$) and 
$\auau$ collisions from STAR~\cite{ref:starauau} ($3 < \pt < 10 \ \gevc$). The bars represent the statistical uncertainty, and the boxes represent the systematic uncertainty.  
The results for $0\--20\%$, $20\--40\%$, and $40\--60\%$ centrality $\auau$ collisions are also shown 
in Fig.~\ref{fig:ptspectrum} (a-c) and are compared to high-$\pt$ data. The STAR data in $\auau$ collisions are consistent with the previously published results from PHENIX~\cite{ref:phenixauau} for $\pt < 5 \ \gevc$ and $|y|<0.35$. 

\begin{figure}[h]
\begin{center}
\includegraphics[width=.5\textwidth]{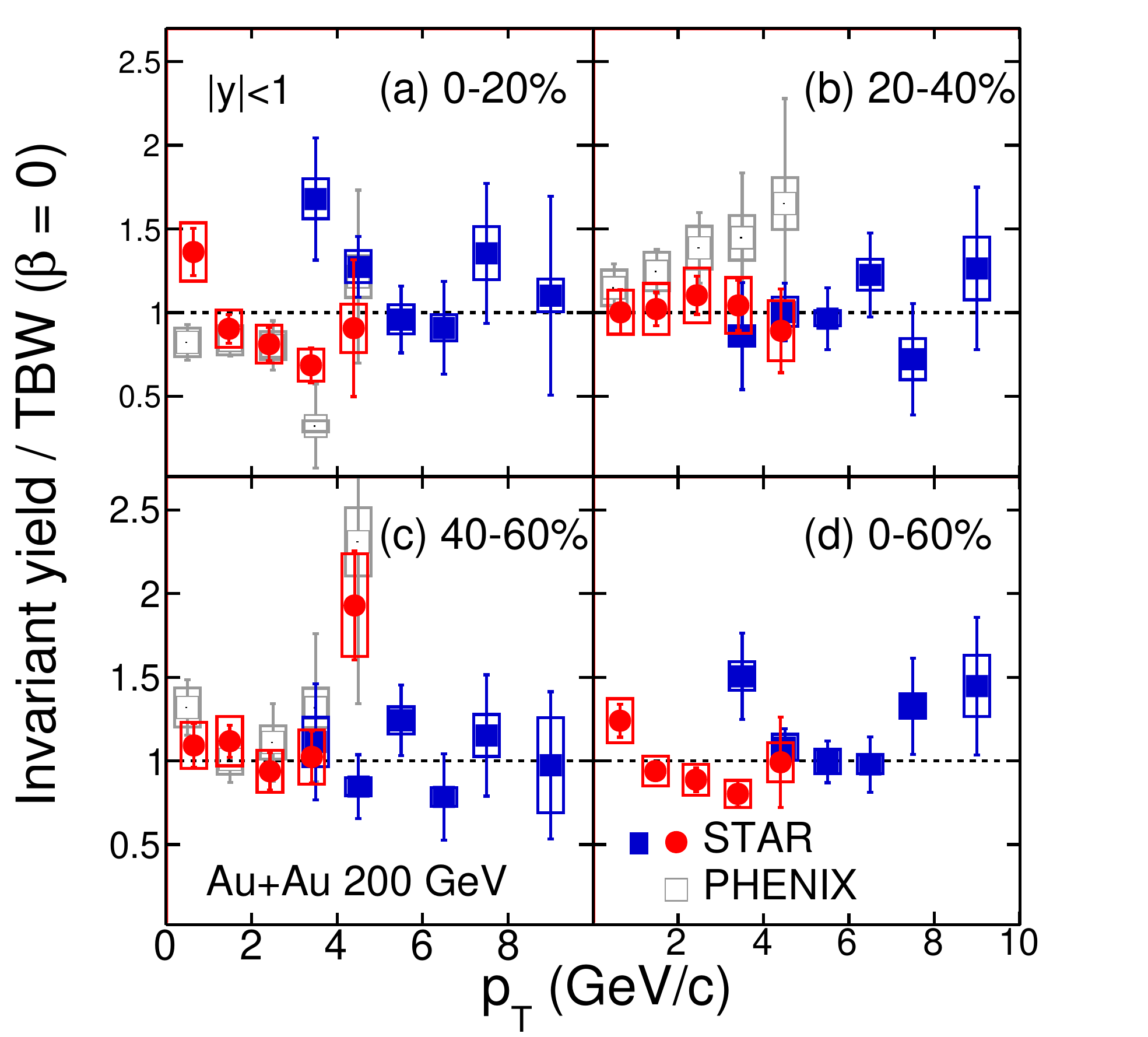}     
\caption{
  (Color online) Ratio of invariant yield from Fig.~\ref{fig:ptspectrum} to predictions of Tsallis blast wave model with radial flow $\beta= 0$~\cite{ref:startbw2}. Data are shown as a function of transverse momentum for $|y|<1$ in (a) $0\--20\%$, (b) $20\--40\%$, (c) $40\--60\%$, and (d) $0\--60\%$ 
    centrality in $\auau$ collisions.   
}
\label{fig:ptspectrumTbwRatio}
\end{center}
\end{figure}

The $\jpsi$ invariant yield in $\auau$ collisions shown in Fig.~\ref{fig:ptspectrum} was compared to a Tsallis blast wave (TBW) model assuming that the $\jpsi$ flows like 
lighter hadrons (dashed line)~\cite{ref:startbw1, ref:startbw2}, i.e. assuming the radial flow velocity $\beta= 0.41$ for $0\--60\%$ central collisions, $0.51$ for $0\--20\%$, $0.44$ for $20\--40\%$ and $0.28$ for $40\--60\%$ events. The normalization for the TBW model was determined from the high $\pt$ data. 
The TBW model qualitatively agrees with our data for $\pt > 2 \ \gevc$: $\chi^2$ over number of degree of freedom (NDF), $\chi^2/{\rm NDF} = 7.2/4$ for  $0\--60\%$ central events, taking into account statistical and systematic uncertainties. However, it underestimates the yield at lower $\pt$:  $\chi^2/{\rm NDF} = 100/9 = 11.1$ for $0 < \pt < 10 \ \gevc$ in  $0\--60\%$ central collisions. The STAR data were also fitted with a TBW model in $0 < \pt < 10 \ \gevc$ that assumes a zero radial flow velocity (solid line)~\cite{ref:startbw1}. Figure~\ref{fig:ptspectrumTbwRatio} shows a ratio of $\jpsi$ invariant yield in Au+Au collisions to the TBW model with $\beta= 0$. The agreement with the data at low $\pt$ is better for semi-central and peripheral events ($\chi^2/{\rm NDF} = 2.9/9$ for $20\--40\%$ and $\chi^2/{\rm NDF} = 13.1/9$ for $40\--60\%$ centrality classes). There is still some discrepancy for central and minimum-bias events: $\chi^2/{\rm NDF} = 26.5/9$ for $0\--20\%$ central events and $\chi^2/{\rm NDF} = 21.7/9$ for $0\--60\%$ events. It suggests, that the $\jpsi$ has a small radial flow, or that there may be contributions from recombination. Recombination is expected to happen at low transverse momenta, thus it would increase the $\jpsi$ yield at low $\pt$.

\begin{figure}[h]
\begin{center}
\includegraphics[width=.5\textwidth]{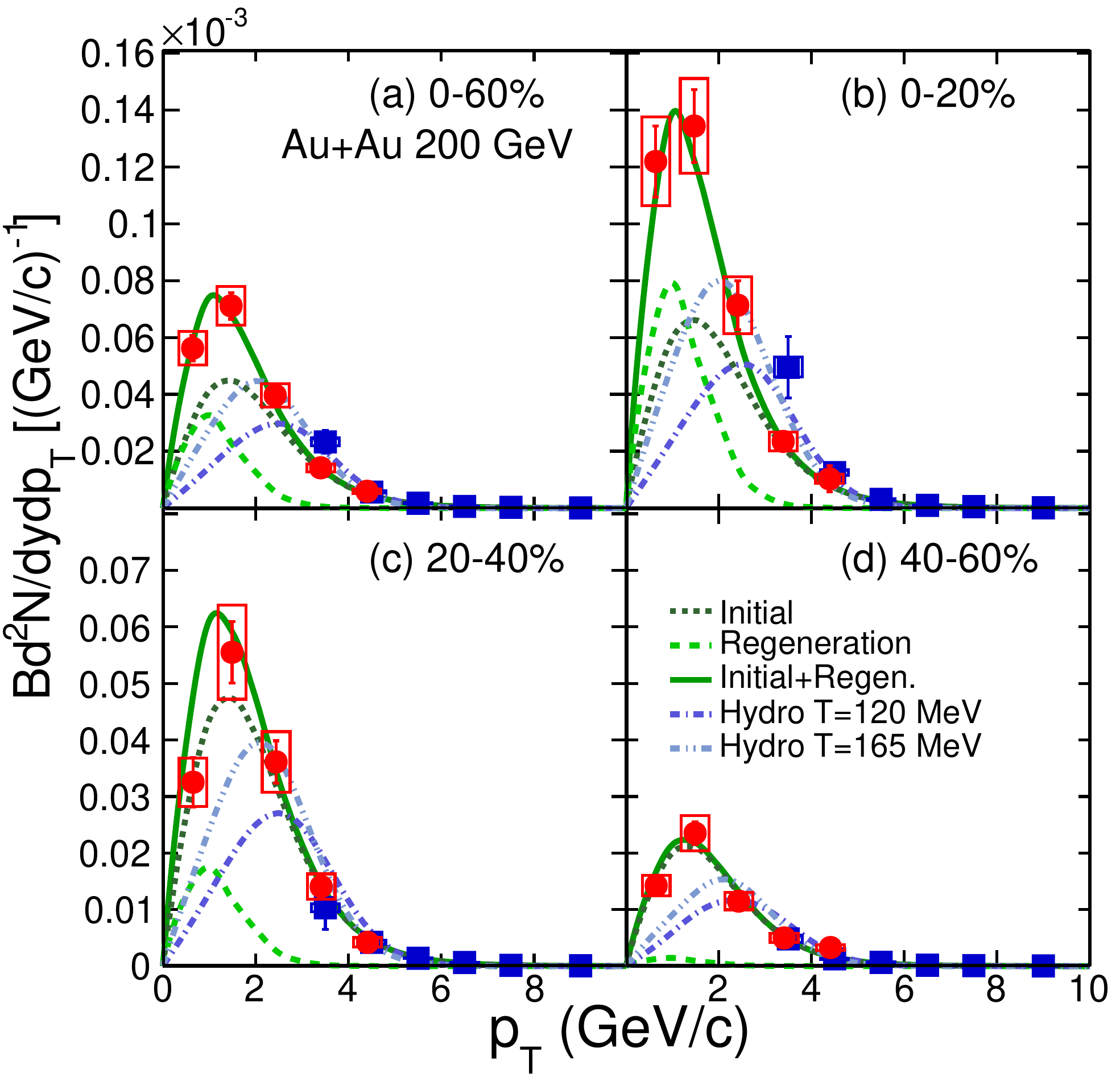}     
\caption{
  (Color online) The $\jpsi$ yield versus transverse momentum for $|y|<1$ in (a) $0\--60\%$, (b) $0\--20\%$, (c) $20\--40\%$, and (d) $40\--60\%$ 
    collision centrality in $\auau$ collisions (solid circles). The data are compared to high-$\pt$ ($3 < \pt < 10 \ \gevc$) results from STAR~\cite{ref:starauau} (solid squares). 
    The models are described in the text~\cite{ref:twocomp2, ref:jpsi_hydro}.
}
\label{fig:dndy}
\end{center}
\end{figure}
The $\jpsi$ yield, $B\mathrm{d}^{2}N/\mathrm{d}y\mathrm{d}\pt$, is shown in Fig.~\ref{fig:dndy} for various collision centralities. The results are compared to predictions from  
viscous hydrodynamics using a $\jpsi$ decoupling temperature of $T = 120 \ \mev$ and $T = 165 \ \mev$ (dot-dashed lines)~\cite{ref:jpsi_hydro}. 
The predictions assume a zero chemical potential for $\jpsi$ at kinetic freeze-out, and the scale of the predictions is determined from a fit to the data in the $\pt$ range of $\pt < 5 \ \gevc$. 
The data favor the higher decoupling temperature; however, the hydrodynamic calculations fail to describe the low $\pt$ $\jpsi$ yield ($\pt < 2 \ \gevc$) and predict a large $\jpsi$ elliptic flow at high-$\pt$, while the measured elliptic flow for $\pt > 2 \ \gevc$ is consistent with zero~\cite{ref:JpsiV2}. 

The data are also compared to theoretical predictions that include $\jpsi$ suppression due to color screening and the statistical regeneration of charm quarks in $\auau$ by Liu et al.~\cite{ref:twocomp2} (dashed line). 
The contribution from initial production dominates in peripheral events. Regeneration becomes more significant in central events and at low $\pt$. 
The predictions describe the $\pt$ spectrum across the entire measured transverse momentum range ($\pt < 10 \ \gevc$). 

The $\jpsi$ yield is summarized in Tables~\ref{tab:yield_pt_auau} and ~\ref{tab:yield_cent_auau} and Tables~\ref{tab:yield_pt_cucu} and ~\ref{tab:yield_cent_cucu} for $\auau$ and $\cucu$ 
collisions, respectively. The uncertainties are separated into (A) statistical and (B) systematic uncertainties. 

\begin{figure*}[hbt!]
\begin{center}
\includegraphics[width=.7\textwidth]{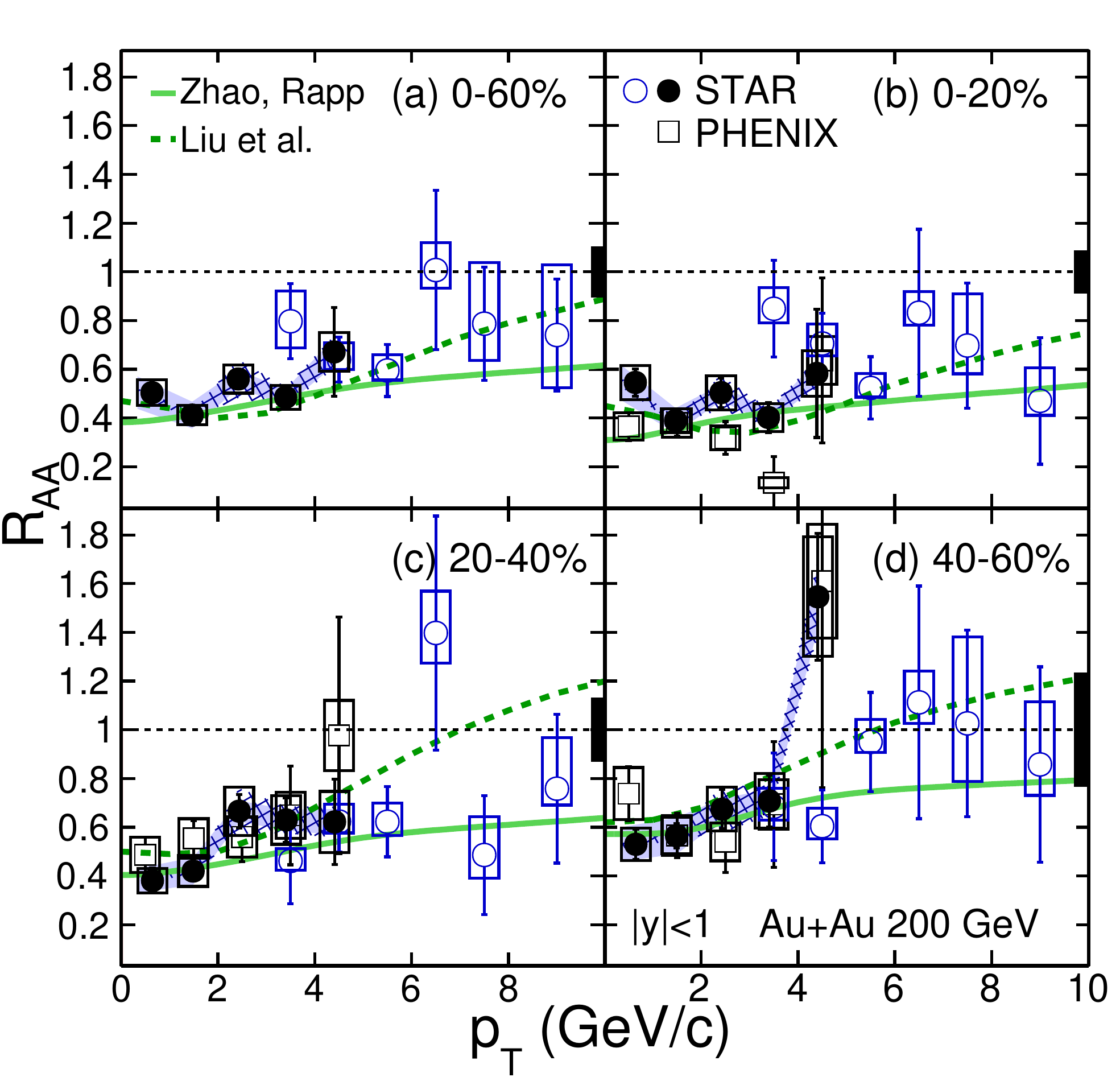}     
\caption{
  (Color online) The $\jpsi$ nuclear modification factor versus transverse momentum for $|y|<1$ and $\pt < 5 \ \gevc$ for 
    (a) $0\--60\%$, (b) $0\--20\%$, (c) $20\--40\%$, and (d) $40\--60\%$
    centrality $\auau$ collisions (solid circles). 
        The data are compared with STAR high-$\pt$ ($5 < \pt < 10 \ \gevc$) results~\cite{ref:starauau} and PHENIX results~\cite{ref:phenixauau} in $|y|<0.35$ (open squares). 
        The statistical and systematic uncertainties on the baseline $\jpsi$ cross section in $\pp$ collisions are indicated by the hatched and solid bands, respectively. Boxes on the vertical axes represent the uncertainty on $\ncoll$ combined with the uncertainty on the inelastic cross section in $\pp$ collisions.
}
\label{fig:raa_pt}
\end{center}
\end{figure*}

\begin{figure}[hbt]
\begin{center}
\includegraphics[width=.5\textwidth]{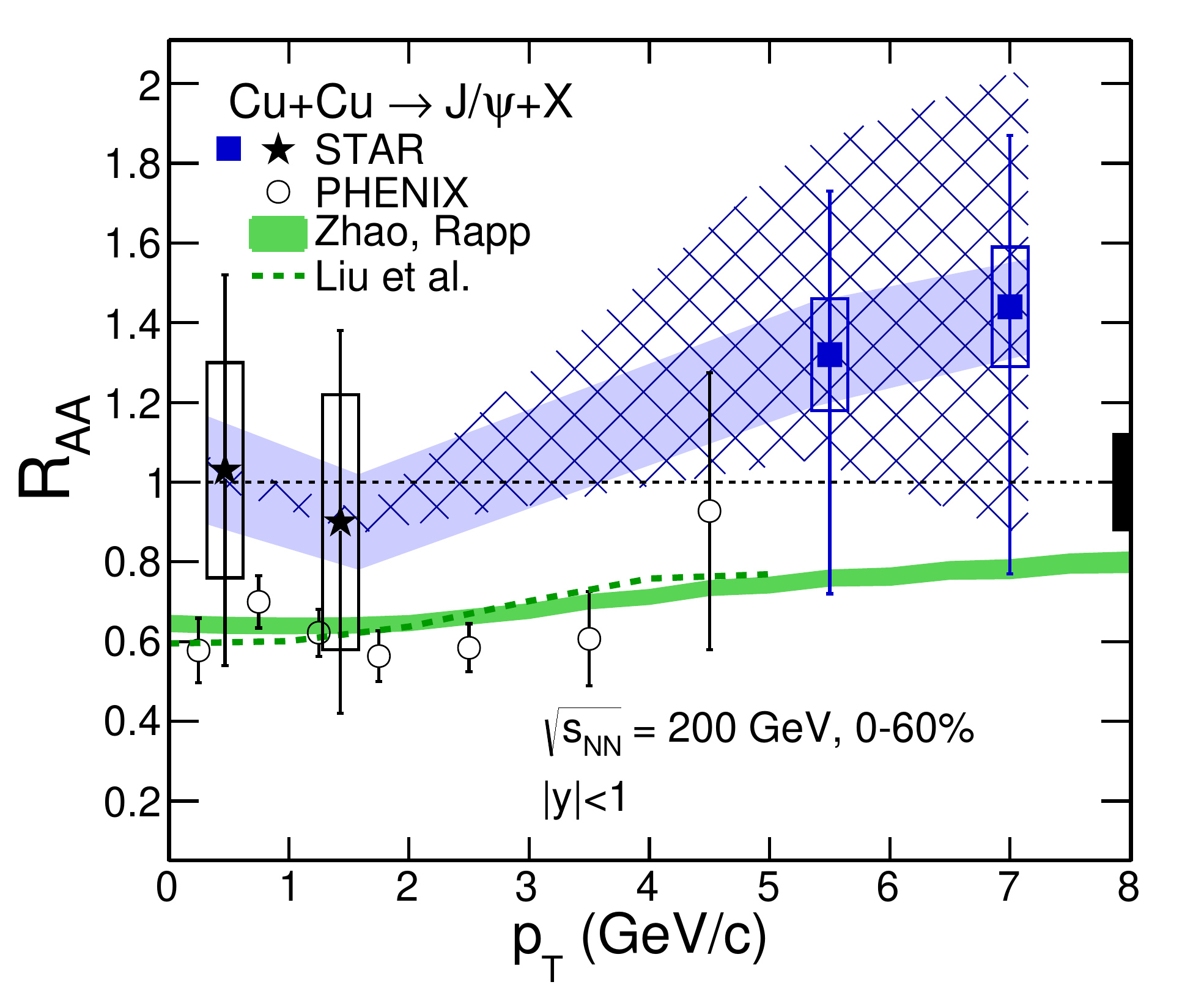}     
\caption{
 	(Color online) The $\jpsi$ nuclear modification factor versus $\pt$ for $|y|<1$ and $\pt < 5 \ \gevc$ in $\cucu$ collisions (solid stars). 
  The data are compared to high-$\pt$ results from STAR (open stars)~\cite{ref:starauau} and 
  PHENIX data with $|y|<0.35$~\cite{ref:phenixcucu} (open circles). The bars and boxes on the data points represent the statistical and systematic uncertainties, respectively. The statistical and systematic uncertainties on the $\jpsi$ cross section in $\pp$ collisions are indicated by the hatched and solid bands, respectively. The box on the vertical axis represents the uncertainty on $\ncoll$ combined with the uncertainty on the inelastic cross section in $\pp$ collisions. 
 }
\label{fig:raa_CuCu}
\end{center}
\end{figure}

\begin{figure}[hbt]
\begin{center}
\includegraphics[width=.5\textwidth]{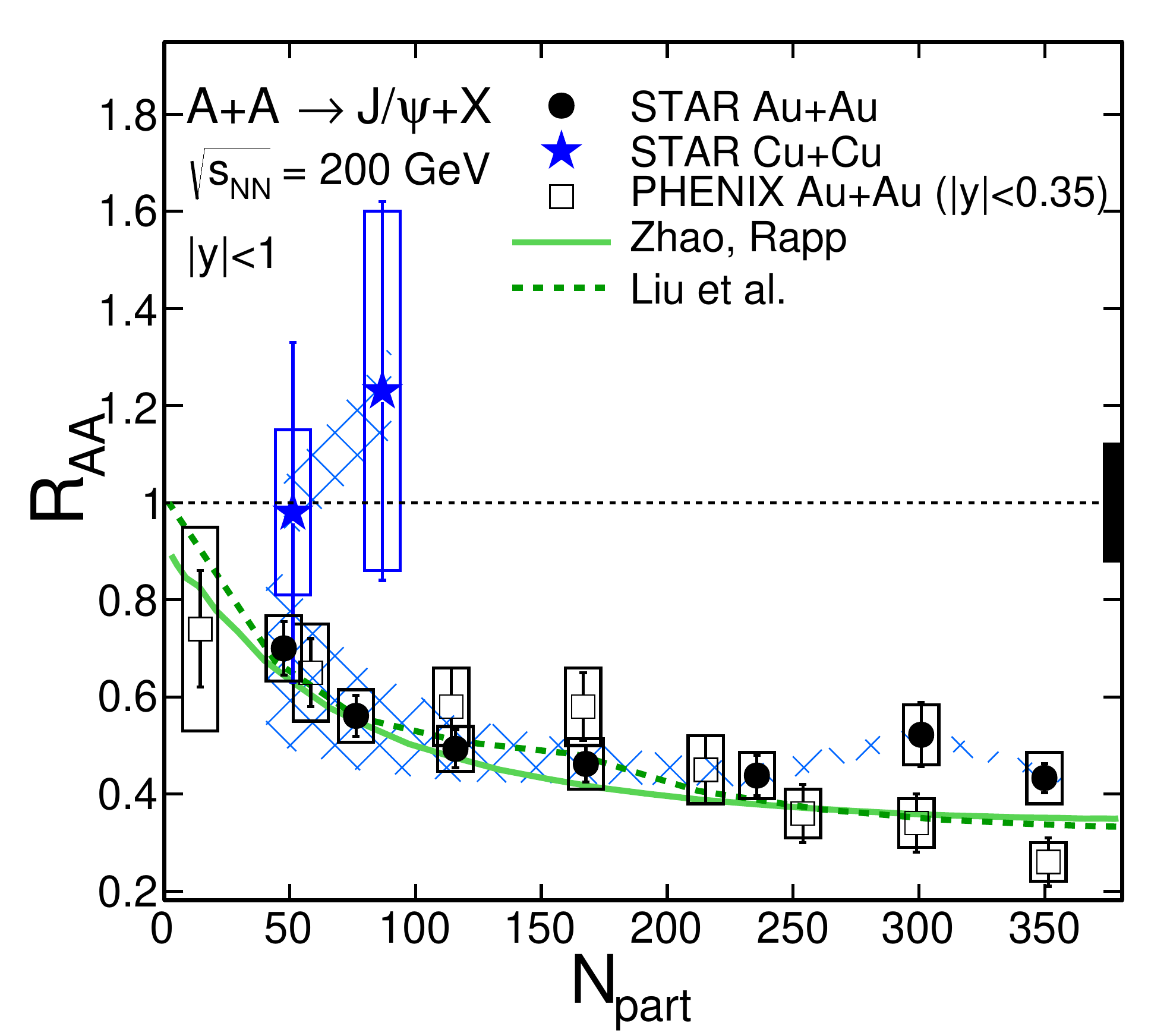}     
\caption{
  (Color online) The nuclear modification factor versus $\npart$ for $\jpsi$ with $|y|<1$ and $\pt < 5 \ \gevc$ in $\auau$ collisions (solid circles) 
    and $\cucu$ collisions (solid stars). 
  The data are compared to PHENIX data with $|y|<0.35$ (open squares) and theoretical predictions (solid line~\cite{ref:twocomp} and dashed line~\cite{ref:twocomp2}). 
  The uncertainty in $\pp$ collisions is described by the box along the vertical axis, 
  and the hatched bands indicate the uncertainty from $\ncoll$.
}
\label{fig:raa_npart}
\end{center}
\end{figure}

To quantify $\jpsi$ suppression in $\AAa$ collisions, we calculate a nuclear modification factor $\raa$. $\raa$ is obtained from the ratio of the $\jpsi$ yield in $\AAa$ and $\pp$ collisions, where the latter is scaled by the average number of binary collisions $\mncoll$ in $\AAa$:
\begin{equation}
\raa = \frac{\sigppinel}{\mncoll}\frac{\mathrm{d}^{2}N_{AA}/\mathrm{d}y\mathrm{d}\pt}{\mathrm{d}^{2}\sigppjpsi/\mathrm{d}y\mathrm{d}\pt}
\end{equation}
where $\sigppinel$ is the inelastic cross section in $\pp$ collisions, $\sigppinel = 42 \pm 3$~mb, $N_{AA}$ is $\jpsi$ yield in $\AAa$ collisions and  
$\mathrm{d}^{2}\sigppjpsi/\mathrm{d}y\mathrm{d}\pt$ is the $\jpsi$ cross section in $\pp$ collisions. 

 The integrated $\jpsi$ cross section in $\pp$ collisions, used as a baseline, was obtained by combining the STAR data for $\pt>2$~GeV/c~\cite{ref:starauau} and low-$\pt$ ($\pt<2$~GeV/c) mid-rapidity measurements from PHENIX~\cite{ref:JpsiPpData}. The global uncertainty combines the statistical and systematic uncertainty on the $\jpsi$ $\pp$ cross section ($\sigppjpsi$)~\cite{ref:JpsiPpData,ref:starauau},  
 the uncertainty on the inelastic cross section in $\pp$ collisions ($\sigppinel$) at STAR ($8\%$)~\cite{ref:norm_uncertain} and PHENIX ($10\%$)~\cite{ref:JpsiPpData}, and the 
 uncertainty in $\ncoll$ shown in Table~\ref{tab:auau_centrality}. The PHENIX results (for $|y|<0.35$) were extrapolated to the STAR acceptance ($|y|<1$) assuming that $\mathrm{d}\sigma /{\mathrm{d}y}$ is constant at $|y|<1$. We estimated the systematic error due to the extrapolation by fitting $\jpsi$ $\mathrm{d}\sigma/{\mathrm{d}y}$ distribution~\cite{ref:Phenix:JpsiRapSpectra} with Gaussian and $A\exp(-b\cosh(cy))$ functions (where $A$, $b$ and $c$ are free parameters), and then calculating the cross-section at $y=0$ using STAR rapidity coverage. We found that this systematic error is about 1\%.
 Figure ~\ref{fig:ptspectrum}(d) shows the combined cross section in $\pp$ collisions with the magnitude indicated by the scale on the right vertical axis. We present here only the PHENIX $\pp$ data points used in the $\raa$ calculations for the sake of clarity; nonetheless STAR and PHENIX results agree very well in the overlapping $\pt$ range~\cite{ref:starauau};

The transverse momentum dependence of the nuclear modification factor is shown in Fig.~\ref{fig:raa_pt} 
for various collision centralities in $\auau$, 
and in Fig.~\ref{fig:raa_CuCu} for $0\--60\%$ centrality $\cucu$ collisions. 
The bars and boxes on the data points represent the statistical and systematic uncertainties, respectively. The statistical and systematic uncertainties on the $\jpsi$ cross section in $\pp$ collisions are indicated by the hatched and solid bands, respectively. The boxes on the vertical axes represent the uncertainty on $\ncoll$, combined with the uncertainty on the inelastic cross section in $\pp$ collisions at STAR of $8\%$~\cite{ref:norm_uncertain}. The $\auau$ and $\cucu$ data are compared to the STAR high-$\pt$ results~\cite{ref:starauau} for $|y|<1$, and to 
PHENIX results~\cite{ref:phenixauau, ref:phenixcucu} in $|y|<0.35$. 
The data in Fig.~\ref{fig:raa_pt} are compared to theoretical predictions based on the suppression of $\jpsi$ due to color screening and 
the statistical regeneration of charm quarks in $\auau$ and $\cucu$ by Zhao and Rapp (solid line~\cite{ref:twocomp,ref:Zhao:Rap:2007hh} and Liu et. all (dashed line)~\cite{ref:twocomp2}). 
The model of Zhao and Rapp also includes $B$ feed-down and formation-time effect (the leakage effect) to $\jpsi$ production. 
Due to limited statistics, STAR results at low $\pt$ are inconclusive regarding possible suppression of $\jpsi$ in Cu+Cu collisions. The $\auau$ data exhibit an increase in $\raa$ for $\pt > 1 \ \gevc$ for all centralities. Both models are able to reproduce the data. A significant suppression is observed for $\pt < 3 \ \gevc$ in $\auau$ collisions ($\raa < 0.6$) for all centralities. 

The centrality dependence of the $\jpsi$ nuclear modification factor, $\raa$, is shown in Fig.~\ref{fig:raa_npart} as a function of $\npart$. The STAR data for $\pt < 5 \ \gevc$ in $\auau$ and $\cucu$ collisions are shown for $|y|<1$. 
The uncertainty on $\ncoll$ in $\auau$ and $\cucu$ is indicated by the hatched point-to-point bands.  The global uncertainty combines  
the uncertainty on the $\jpsi$ cross section in $\pp$ collisions and the uncertainty on the inelastic cross section in $\pp$ collisions at STAR, and is 
indicated by the band on the right vertical axis.

PHENIX previously reported a significant suppression in mid-central and central $\cucu$ collisions~\cite{ref:phenixcucu}. The STAR $\cucu$ data exhibit no suppression within sizable uncertainties. However, the difference between STAR and PHENIX results is less than $1.5$ standard deviation when systematic and statistical uncertainties are taken into account.

The $\auau$ data are suppressed for all centralities. The suppression increases with collision centrality up to $\npart \sim 150$ and then saturates. The data are compared to the PHENIX results in $\auau$ collisions with $|y|<0.35$~\cite{ref:phenixauau}. These results are consistent for peripheral and semi-central collisions ($\npart < 250$). For the most central collisions (0-5\% and 0-10\%) the STAR data show a smaller suppression compared to PHENIX results. Nevertheless, the difference between these measurements is not statistically significant taking into account statistical and systematic uncertainties: We test the consistency between these results using the z-test: $z = (\mu_1 - \mu_2)/ \sqrt{\sigma^{2}_1 + \sigma^{2}_2}$, where $\mu$ and $\sigma$ is a mean and standard deviation of a given sample, $\sigma = \sqrt{\sigma_{\rm stat.}^2 + \sigma_{\rm syst.}^2}$, and the two samples are assumed to be independent of one another and have a normal distribution. The difference between STAR and PHENIX results for $0-5\%$ most central events is 2$\sigma$ and 1.5$\sigma$ for $5-10\%$ central events.  

The centrality dependence of $\raa$ for $\pt < 5 \ \gevc$ from theoretical predictions for $\jpsi$ production including the suppression of $\jpsi$ due to color screening and 
the statistical regeneration of charm quarks in $\auau$ collisions from Zhao and Rapp~~\cite{ref:twocomp} and Liu et. all~\cite{ref:twocomp2} are also shown.  
    The predictions are able to describe the data well across the collision centrality range, and we are  unable to distinguish between the models.
The $\jpsi$ nuclear modification factor is summarized in Tables~\ref{tab:yield_pt_auau} and ~\ref{tab:yield_cent_auau} for Au+Au collisions and Tables~\ref{tab:yield_pt_cucu} and ~\ref{tab:yield_cent_cucu} for $\cucu$ collisions. The uncertainties are separated into (A) statistical, (B) systematic, and (C) global uncertainties. 

\clearpage

\section{Summary\label{summary}}

We presented $\jpsi$ production at low $\pt$ ($\pt < 5 \ \gevc$) in $\auau$ and $\cucu$ collisions at $\snn = 200 \ \gev$.  These results, combined with STAR high-$\pt$ data, provides coverage of $\jpsi$ production in $\auau$ collisions for a wide $\pt$ range of $0 < \pt < 10 \ \gevc$. Comparisons of $\pt$ spectra with Tsallis blast wave model suggest that the $\jpsi$ has a small radial flow, or significant contribution from recombination at low $\pt$. 

In the case of $\auau$, we observed a strong suppression at low and moderate $\pt$ ($\pt < 3 \ \gevc$) with $\raa < 0.6$ for all centralities. The suppression decreases with increasing $\pt$ for $\pt > 2 \ \gevc$. Measurement of the nuclear modification factor as a function of centrality indicates a strong suppression in central and semi-central collisions. Centrality and $\pt$ dependence of $\raa$ are well described by the models assuming an interplay between color screening and regeneration in the hot medium, as well as possible $\jpsi$ escape effects. A detailed investigation of low-$\pt$ production, including a high-statistics analysis of elliptic flow, may provide a better understanding of this process. 

We thank the RHIC Operations Group and RCF at BNL, the NERSC Center at LBNL, the KISTI Center in Korea, and the Open Science Grid consortium for providing resources and support. This work was supported in part by the Offices of NP and HEP within the U.S. DOE Office of Science, the U.S. NSF, CNRS/IN2P3, FAPESP CNPq of Brazil,  the Ministry of Education and Science of the Russian Federation, NNSFC, CAS, MoST and MoE of China, the Korean Research Foundation, GA and MSMT of the Czech Republic, FIAS of Germany, DAE, DST, and CSIR of India, the National Science Centre of Poland, National Research Foundation (NRF-2012004024), the Ministry of Science, Education and Sports of the Republic of Croatia, and RosAtom of Russia.
\bibliography{Bibliography.bib}{}
\bibliographystyle{apsrev}

\begin {table*} [tbp]
\begin {center}
\caption{The $\jpsi$ invariant yield $\frac{B}{2\pi p_T}\frac{\mathrm{d}^2N}{\mathrm{d}y\mathrm{d}p_T}$ and nuclear modification factor as a function of transverse momentum for $|y|<1$ in $\auau$ collisions at $\snn = 200 \ \gev$ 
  with (A)~statistical, (B)~systematic, and (C)~global uncertainties. The yield and corresponding uncertainties are in units of $(\gevc)^{-2}$. }
\label{tab:yield_pt_auau}
\begin {ruledtabular}
\begin {tabular} { l l l l l l l l l l l l }
\noalign{\smallskip}
 Centrality & $ \pt (\gevc) $ & $\langle \pt \rangle (\gevc) $ &  $\frac{B}{2\pi p_T}\frac{\mathrm{d}^2N}{\mathrm{d}y\mathrm{d}p_T}$ & (A) & +(B) & -(B) & $ \raa $ & (A) & +(B) & -(B) & (C) \\

\noalign{\smallskip}\hline\noalign{\smallskip}

$ 0-60 $ & $ 0-1 $ & $ 0.64$  & $ 17.89\times10^{-6} $ & $ 1.42\times10^{-6} $ & $ +1.89\times10^{-6} $ & $ -1.90\times10^{-6} $ & $ 0.50 $ & $ 0.04 $ & $ +0.05 $ & $ -0.05 $ & $ 0.09 $\\
$ 0-60 $ & $ 1-2 $ & $ 1.47$  & $ 7.54\times10^{-6} $ & $ 0.50\times10^{-6} $ & $ +0.72\times10^{-6} $ & $ -0.73\times10^{-6} $ & $ 0.41 $ & $ 0.03 $ & $ +0.04 $ & $ -0.04 $ & $ 0.08 $\\
$ 0-60 $ & $ 2-3 $ & $ 2.42 $  & $ 2.52\times10^{-6} $ & $ 0.20\times10^{-6} $ & $ +0.26\times10^{-6} $ & $ -0.26\times10^{-6} $ & $ 0.56 $ & $ 0.04 $ & $ +0.06 $ & $ -0.06 $ & $ 0.11 $\\
$ 0-60 $ & $ 3-4 $ & $ 3.40$  & $ 0.64\times10^{-6} $ & $ 0.06\times10^{-6} $ & $ +0.06\times10^{-6} $ & $ -0.06\times10^{-6} $ & $ 0.49 $ & $ 0.05 $ & $ +0.05 $ & $ -0.05 $ & $ 0.08 $\\
$ 0-60 $ & $ 4-5 $ & $ 4.40$  & $ 0.21\times10^{-6} $ & $ 0.06\times10^{-6} $ & $ +0.02\times10^{-6} $ & $ -0.02\times10^{-6} $ & $ 0.67 $ & $ 0.18 $ & $ +0.08 $ & $ -0.08 $ & $ 0.12 $\\
\noalign{\smallskip}
$ 0-20 $ & $ 0-1 $ & $ 0.64$  & $ 38.78\times10^{-6} $ & $ 3.99\times10^{-6} $ & $ +4.96\times10^{-6} $ & $ -4.97\times10^{-6} $ & $ 0.55 $ & $ 0.06 $ & $ +0.07 $ & $ -0.07 $ & $ 0.09 $\\
$ 0-20 $ & $ 1-2 $ & $ 1.47$  & $ 14.25\times10^{-6} $ & $ 1.35\times10^{-6} $ & $ +1.77\times10^{-6} $ & $ -1.77\times10^{-6} $ & $ 0.39 $ & $ 0.04 $ & $ +0.05 $ & $ -0.05 $ & $ 0.07 $\\
$ 0-20 $ & $ 2-3 $ & $ 2.41$  & $ 4.54\times10^{-6} $ & $ 0.55\times10^{-6} $ & $ +0.63\times10^{-6} $ & $ -0.63\times10^{-6} $ & $ 0.50 $ & $ 0.06 $ & $ +0.07 $ & $ -0.07 $ & $ 0.09 $\\
$ 0-20 $ & $ 3-4 $ & $ 3.39$  & $ 1.07\times10^{-6} $ & $ 0.16\times10^{-6} $ & $ +0.15\times10^{-6} $ & $ -0.15\times10^{-6} $ & $ 0.40 $ & $ 0.06 $ & $ +0.06 $ & $ -0.06 $ & $ 0.06 $\\
$ 0-20 $ & $ 4-5 $ & $ 4.39$  & $ 0.36\times10^{-6} $ & $ 0.16\times10^{-6} $ & $ +0.06\times10^{-6} $ & $ -0.06\times10^{-6} $ & $ 0.58 $ & $ 0.26 $ & $ +0.09 $ & $ -0.09 $ & $ 0.10 $\\
\noalign{\smallskip}
$ 20-40 $ & $ 0-1 $ & $ 0.65 $  & $ 10.35\times10^{-6} $ & $ 1.38\times10^{-6} $ & $ +1.35\times10^{-6} $ & $ -1.36\times10^{-6} $ & $ 0.38 $ & $ 0.05 $ & $ +0.05 $ & $ -0.05 $ & $ 0.08 $\\
$ 20-40 $ & $ 1-2 $ & $ 1.49 $  & $ 5.89\times10^{-6}$ & $ 0.58\times10^{-6}$ & $ +0.88\times10^{-6}$ & $ -0.88\times10^{-6}$ & $ 0.42 $ & $ 0.04 $ & $ +0.06 $ & $ -0.06 $ & $ 0.08 $\\
$ 20-40 $ & $ 2-3 $ & $ 2.44 $  & $ 2.30\times10^{-6}$ & $ 0.24\times10^{-6}$ & $ +0.34\times10^{-6}$ & $ -0.34\times10^{-6}$ & $ 0.67 $ & $ 0.07 $ & $ +0.10 $ & $ -0.10 $ & $ 0.14 $\\
$ 20-40 $ & $ 3-4 $ & $ 3.41 $  & $ 0.64\times10^{-6}$ & $ 0.09\times10^{-6}$ & $ +0.10\times10^{-6}$ & $ -0.10\times10^{-6}$ & $ 0.63 $ & $ 0.09 $ & $ +0.10 $ & $ -0.10 $ & $ 0.11 $\\
$ 20-40 $ & $ 4-5 $ & $ 4.41 $  & $ 0.15\times10^{-6} $ & $ 0.04\times10^{-6} $ & $ +0.03\times10^{-6} $ & $ -0.03\times10^{-6} $ & $ 0.62 $ & $ 0.17 $ & $ +0.13 $ & $ -0.13 $ & $ 0.12 $\\
\noalign{\smallskip}
$ 40-60 $ & $ 0-1 $ & $ 0.65 $  & $ 4.53\times10^{-6} $ & $0.54\times10^{-6}$ & $ +0.57\times10^{-6} $ & $ -0.57\times10^{-6} $ & $ 0.53 $ & $ 0.06 $ & $ +0.07 $ & $ -0.07 $ & $ 0.15 $\\
$ 40-60 $ & $ 1-2 $ & $ 1.49 $  & $ 2.49\times10^{-6} $ & $ 0.22\times10^{-6} $ & $ +0.33\times10^{-6}$ & $ -0.33\times10^{-6}$ & $ 0.57 $ & $ 0.05 $ & $ +0.07 $ & $ -0.07 $ & $ 0.16 $\\
$ 40-60 $ & $ 2-3 $ & $ 2.43 $  & $ 0.73\times10^{-6}$ & $ 0.09\times10^{-6}$ & $ +0.10\times10^{-6}$ & $ -0.10\times10^{-6}$ & $ 0.67 $ & $ 0.08 $ & $ +0.09 $ & $ -0.09 $ & $ 0.19 $\\
$ 40-60 $ & $ 3-4 $ & $ 3.41 $  & $ 0.23\times10^{-6}$ & $ 0.03\times10^{-6}$ & $ +0.04\times10^{-6}$ & $ -0.04\times10^{-6} $ & $ 0.71 $ & $ 0.11 $ & $ +0.11 $ & $ -0.11 $ & $ 0.19 $\\
$ 40-60 $ & $ 4-5 $ & $ 4.41 $  & $ 0.12\times10^{-6} $ & $ 0.02\times10^{-6} $ & $ +0.02\times10^{-6} $ & $ -0.02\times10^{-6} $ & $ 1.55 $ & $ 0.26 $ & $ +0.24 $ & $ -0.25 $ & $ 0.43 $\\
\end {tabular}
\end {ruledtabular}
\end {center}
\end {table*}

\begin {table*} [tbp]
\begin {center}
\caption{The $\jpsi$ invariant yield and nuclear modification factor as a function of centrality for $|y|<1$ in $\auau$ collisions at $\snn = 200 \ \gev$ 
  with (A)~statistical, (B)~systematic, and (C)~global uncertainties. }
\label{tab:yield_cent_auau}
\begin {ruledtabular}
\begin {tabular} { l l l l l l l l l l l }
\noalign{\smallskip}
 Centrality & $ \pt \ (\gevc)$ & ${B}{\mathrm{d}N}/{\mathrm{d}y}$ & (A) & +(B) & -(B) & $ \raa $ & (A) & +(B) & -(B) & (C) \\

\noalign{\smallskip}\hline\noalign{\smallskip}
$ 0-5 $ & $ 0-5 $ & $ 464.38\times10^{-6} $ & $ 32.43\times10^{-6} $ & $ +56.67\times10^{-6} $ & $ -56.82\times10^{-6} $ & $ 0.43 $ & $ 0.03 $ & $ +0.05 $ & $ -0.05 $ & $ 0.06 $\\
$ 5-10 $ & $ 0-5 $ & $ 447.06\times10^{-6} $ & $ 56.24\times10^{-6} $ & $ +52.95\times10^{-6} $ & $ -53.10\times10^{-6} $ & $ 0.52 $ & $ 0.07 $ & $ +0.06 $ & $ -0.06 $ & $ 0.07 $\\
$ 10-20 $ & $ 0-5 $ & $ 266.75\times10^{-6} $ & $ 25.70\times10^{-6} $ & $ +29.06\times10^{-6} $ & $ -29.13\times10^{-6} $ & $ 0.44 $ & $ 0.04 $ & $ +0.05 $ & $ -0.05 $ & $ 0.06 $\\
$ 20-30 $ & $ 0-5 $ & $ 174.38\times10^{-6} $ & $ 14.01\times10^{-6} $ & $ +19.50\times10^{-6} $ & $ -19.55\times10^{-6} $ & $ 0.46 $ & $ 0.04 $ & $ +0.05 $ & $ -0.05 $ & $ 0.07 $\\
$ 30-40 $ & $ 0-5 $ & $ 110.44\times10^{-6} $ & $ 8.77\times10^{-6} $ & $ +10.26\times10^{-6} $ & $ -10.28\times10^{-6} $ & $ 0.49 $ & $ 0.04 $ & $ +0.05 $ & $ -0.05 $ & $ 0.09 $\\
$ 40-50 $ & $ 0-5 $ & $ 69.85\times10^{-6} $ & $ 5.19\times10^{-6} $ & $ +6.74\times10^{-6} $ & $ -6.76\times10^{-6} $ & $ 0.56 $ & $ 0.04 $ & $ +0.05 $ & $ -0.05 $ & $ 0.13 $\\
$ 50-60 $ & $ 0-5 $ & $ 45.15\times10^{-6} $ & $ 3.52\times10^{-6} $ & $ +4.30\times10^{-6} $ & $ -4.32\times10^{-6} $ & $ 0.70 $ & $ 0.05 $ & $ +0.07 $ & $ -0.07 $ & $ 0.21 $\\

\end {tabular}
\end {ruledtabular}
\end {center}
\end {table*}

\begin {table*} [tbp]
\begin {center}
\caption{The $\jpsi$ invariant yield $\frac{B}{2\pi p_T}\frac{\mathrm{d}^2N}{\mathrm{d}y\mathrm{d} p_T}$ and nuclear modification factor as a function of transverse momentum for $|y|<1$ in $\cucu$ collisions at $\snn = 200 \ \gev$ 
  with (A)~statistical, (B)~systematic, and (C)~global uncertainties. The yield and corresponding uncertainties are in units of $(\gevc)^{-2}$. }
\label{tab:yield_pt_cucu}
\begin {ruledtabular}
\begin {tabular} { l l l l l l l l l l l l }
\noalign{\smallskip}
 Centrality & $ \pt (\gevc) $ & $\langle \pt \rangle (\gevc) $ & $\frac{B}{2\pi p_T}\frac{\mathrm{d}^2N}{\mathrm{d}y\mathrm{d} p_T}$ & (A) & +(B) & -(B) & $ \raa $ & (A) & +(B) & -(B) & (C) \\
\noalign{\smallskip}\hline\noalign{\smallskip}

 $ 0-60 $ & $ 0-1 $ & $0.47$ & $7.48\times10^{-6} $ & $3.57\times10^{-6}$ & $2.0\times10^{-6}$ & $2.0\times10^{-6}$ & $ 1.03 $ & $ 0.49 $ & $ 0.27 $ & $ 0.27 $ & $ 0.19 $\\ 
          & $ 1-2 $ & $1.43$ & $3.56\times10^{-6}$ & $1.88\times10^{-6}$ & $1.27\times10^{-6}$ & $1.27\times10^{-6}$ &  $ 0.95 $ & $ 0.50 $ & $ 0.34 $ & $ 0.34 $ & $ 0.17 $\\
\end {tabular}
\end {ruledtabular}
\end {center}
\end {table*}

\begin {table*} [tbp]
\begin {center}
\caption{The $\jpsi$ invariant yield and nuclear modification factor for $|y|<1$ in $\cucu$ collisions at $\snn = 200 \ \gev$ 
  with (A)~statistical, (B)~systematic, and (C)~global uncertainties.}
\label{tab:yield_cent_cucu}
\begin {ruledtabular}
\begin {tabular} { l l l l l l l l l l l }
\noalign{\smallskip}
 Centrality & $ \pt (\gevc) $ &  ${B}{\mathrm{d}N}/{\mathrm{d}y}$ & (A) & +(B) & -(B) & $ \raa $ & (A) & +(B) & -(B) & (C) \\
\noalign{\smallskip}\hline\noalign{\smallskip}

 $ 0-20 $ & $ 0-5 $ & $ 192\times10^{-6} $ & $ 61\times10^{-6} $ & $ 59\times10^{-6} $ & $ 59\times10^{-6} $ & $ 1.23 $ & $ 0.39 $ & $ 0.37 $ & $ 0.37 $ & $ 0.15 $ \\
 $ 0-60 $ & $ 0-5 $ & $ 79\times10^{-6} $ & $ 22\times10^{-6} $ & $ 11\times10^{-6} $ & $11\times10^{-6}$ & $ 0.98 $ & $0.35$ & $0.17$ & $0.17$ & $0.15$  \\

\end {tabular}
\end {ruledtabular}
\end {center}
\end {table*}

\clearpage

\end{document}